%% file: finalspinnet.tex
\documentstyle[epsfig,aps,amssymb,amsfonts,epic,eepic,preprint,tighten]{revtex}
\makeatletter
\@addtoreset{equation}{section}
\renewcommand{\theequation}{\thesection.\arabic{equation}}

\renewcommand{\thesection}{\arabic{section}}
\renewcommand{\thesubsection}{\thesection.\arabic{subsection}}
\def\ni{\noindent}

\newcommand{\Ref}[1]{(\ref{#1})}
\def\tr{\textrm}
\def\subs{\subsection}
\def\sss{\subsubsection}

\def\Rnum{{\bf R}}

\newcommand{\Z}{{\bf Z}}
\newcommand{\C}{{\bf C}}

\let\pp=\partial

\def\beq{\begin{equation}}
\def\eeq{\end{equation}}
\def\beqs{\begin{eqnarray}}
\def\eeqs{\end{eqnarray}}
\def\beqsl{\begin{eqnarray}{l}}
\def\eeqs{\end{eqnarray}}

\def\lrarr{\leftrightarrow}
\def\arr{\rightarrow}
\def\dmu{\tr{d}\mu}
\def\tl{\tilde}
\def\wtl {\widetilde}
\def\G{{\cal G}}
\def\Gtl{\wtl{\G}}
\def\H{{\cal H}}
\def\Htl{\wtl{\H}}
\def\gt{G^{(T)}}
\def\gu{G^{(U)}}
\def\bA{\bar{A}}
\def\om{\omega}

\def\su{{\rm SU}(2)}
\def\slc{{\rm SL}(2,\C)}
\def\slr{{\rm SL}(2,\R)}

\def\A{{\cal{A}}}
\def\G{{\cal{G}}}

\def\R{{\bf R}}

\def\sc#1#2{<#1|#2>}

\newcommand{\mat}[4]{
\left( \begin{array}{cc} #1 &#2\\ #3 &#4\end{array} \right)
}
\def \f{\frac}
\def \sh{\sinh}

\def\PBOX[#1]#2{\mbox{\setlength{\unitlength}{#1 pt} #2}}
\def\CBOX[#1]#2{\begin{array}{c} \PBOX[#1]{#2} \end{array}}
\def\MPIC[#1,#2]#3{{\begin{picture}(#1,#2) {#3} \end{picture}}}
\def\FIGThetaNet#1#2#3{{\MPIC[40,40]{{
        \put(18,32){$#1$}
        \put( 0,15){\line(1,0){40}} \put(18,17){$#2$}
        \put(16,15){\oval(40,30)}   \put(18, 2){$#3$}
        \put( 0,15){\circle*{3}}    \put(40,15){\circle*{3}}
}}}}
\def\looptwo#1#2{{\MPIC[40,40]{{
        \put(18,32){$#1$}
        \put(16,15){\oval(40,30)}   \put(18, 2){$#2$}
        \put( 0,15){\circle*{3}}    \put(40,15){\circle*{3}}
}}}}
\def\loopone#1{{\MPIC[40,40]{{
        \put(18,32){$#1$}
        \put(16,15){\oval(40,30)}
        \put( 0,15){\circle*{3}}
}}}}

\newtheorem{theo}{Theorem}
\newtheorem{prop}{Proposition}
\newtheorem{defi}{Definition}

\begin{document}

\title{Spin Networks for Non-Compact Groups}
\author{
{\bf Laurent Freidel}
\thanks{e.mail:freidel@ens-lyon.fr}
}
\address{
Perimeter Institute for Theoretical Physics\\
35 King street North, Waterloo  N2J-2G9,Ontario, Canada\\
and \\
Laboratoire de Physique, \'Ecole Normale Sup{\'e}rieure de Lyon \\
46 all{\'e}e d'Italie, 69364 Lyon Cedex 07, France\thanks{UMR 5672 du CNRS}
}
\author{
{\bf Etera R. Livine}\thanks{e.mail: livine@cpt.univ-mrs.fr}
}
\address{
Centre de Physique Th{\'e}orique,
 Campus de Luminy, Case 907\\
13288 Marseille Cedex 09, France\thanks{Unit\'e propre 7061 du CNRS}\\
}

\date{\normalsize \today}
\maketitle

\begin{abstract}
Spin networks are a natural generalization of Wilson loop
functionals. They have been extensively studied in the case where
the gauge group is compact and it has been shown that they
naturally form a basis of gauge invariant observables. Physically
the restriction to compact gauge groups is enough for the study of
Yang-mills theories, however it is well known that non-compact
groups naturally arise as internal gauge groups for Lorentzian
gravity models. In this context, a proper construction of gauge
invariant observables is needed. The purpose of the present work is to
define the notion of spin network states for non-compact
groups. We first build, by a careful gauge fixing procedure,
a natural measure and a Hilbert space structure on the space of gauge
invariant graph connections. Spin networks are then defined as generalized
eigenvectors of a complete set of hermitic commuting operators.
We show how the delicate issue of taking the quotient of a space by non compact groups
can be address in term of algebraic geometry.
We finally construct the full Hilbert space containing all spin network states.
Having in mind applications to gravity,
we illustrate our results for the groups $\slr$ and $\slc$.
\end{abstract}

%\newpage

%\tableofcontents

\newpage

\section{Introduction}

The purpose of this paper is to generalize the construction of an
Hilbert space of Spin Network states to the case in which the
gauge group is non-compact. Spin network states arise naturally in
many fields of physics since they form a basis of gauge invariant
functionals in Yang-Mills like theories \cite{Kogsuss}.
In the context of gravity
they were introduced by Rovelli and Smolin \cite{rs1,rs} and they were
promoted as a basis of an Hilbert space of gauge and
diffeomorphism invariant functionals in works by Ashtekar and
Lewandowski \cite{al1,all,al2} and by Baez \cite{baez}. This
series of works have been focused on the case where the gauge group is
compact. Compact gauge groups are natural as symmetry groups of
gauge theory and Euclidean gravity, however non-compact gauge
groups arise as symmetry groups of Lorentzian gravity.

For
instance, $SL(2,\C)$ arises in the original Ashtekar formulation
of $3+1$ gravity in terms of self-dual variables. However, in this
context, the lack of properly well defined spin network states has
forced the community to work with the real $SU(2)$ Barbero
\cite{barbero} connection at the price of introducing a new
constant (Immirzi parameter), a more complicated dynamics
\cite{thiemann} and the loss of a natural 4-dimensional
geometrical interpretation of the phase space variables
\cite{samuel}. In the case of $2+1$ Lorentzian gravity, the
partition function and the transition amplitudes have been
computed in terms of spin networks (recoupling coefficients) of
$SL(2,\R)$ \cite{laurent}. In this context, geometrical
interpretation of the representation labels has led to the
conclusion that space is continuous whereas time is discrete
\cite{laurent,crl}, in agreement with former results from 't~Hooft
\cite{thooft}.

In order to put these results on a firmer basis and
construct geometrical operators in the Lorentzian context, one
needs to understand better the nature of non-compact spin networks
and how they form a natural Hilbert space related to the Hilbert
space of gauge and diffeomorphism invariant connections. This is
the purpose of this work.

Several issue concerning non-compact spin network have already been
raised in the literature.
First, Ashtekar and Lewandowski have already addressed the issue of completeness
of the spin network functionals \cite{al3} versus the separability of the space
of gauge invariant connections. We shall come back to these issues in
section \ref{quotient}.
There also has been some attempts to define non compact spin
networks. Marolf \cite{marolf} was the first one to show, in the context of 2+1
gravity on the torus, that the loop transform using finite
dimensional representation is ill defined when the group is non
compact. He and, Ashtekar and Loll \cite{aloll}, have then studied the possibility to
overcome this difficulty at the price of introducing additional and
non natural structures.

Our approach shed new lights on this problem, it leads to a  different point of view
since we don't insist on  having spin networks labelled by finite
dimensional representations. We show that one should work
instead with the infinite dimensional unitary representations of the
group.
The emergence of spin networks labelled by infinite dimensional
representation is not new. They already appeared in the context of spin foams
models for Lorentzian 3d gravity \cite{laurent,davids} and 4-d gravity
\cite{barcra,rovperez}.

Moreover our formalism is
more general since  it sets a framework for all non-compact groups.
Indeed, our purpose is to give a general account of the construction of
non-compact spin networks and the structure of the stratified space
of gauge invariant connections. Our presentation is valid for any
semi-simple reductive group. The general exposition is therefore quite
mathematical. Having in mind further application to gravity
we will  illustrate the main
problems and results in the context of $\slc$ and $\slr$.

\subsection{Connection space and Cylindrical functions}

We choose once for all a manifold $\Sigma$ and $P$ a locally
trivial smooth principal $G$-bundle over $\Sigma$, with $G$ a
semi-simple reductive group. We denote by $\A$ the space of
gauge $G$-connections and by $\G$ the gauge group acting on
connection by $A^k=k^{-1}Ak+k^{-1}dk$. The theories we are
interested in are Yang-Mills like in the sense that the phase
space conjugate variables are given by a G-connection $A$ (a
magnetic potential) and a $adP$ valued densitised vector field
$E$, both are anti-hermitian. This phase space is the cotangent
bundle to the space of connection $T^*(\A)$.
On such a phase space we want to impose the Gauss laws (gauge
invariance) and eventually the diffeomorphism constraint. The
representation of the operator algebra is done in the polarization
where the wave functionals depend on the connection and therefore
the Hilbert space structure is formally $L^2(\A/\G, d\mu)$. The
purpose of this work is to study the structure of this space. In
fact we will first restrict our intention to special  gauge
invariant functionals of the connections  called cylindrical
functionals and we will study the possibility to give the space of
cylindrical functions an  Hilbert space structure.
 In the gauge invariant context, cylindrical
functions are associated with graphs. Given a smooth {\it oriented}
graph $\Gamma$ composed of $E$ oriented edges and $V$ vertices we
have the holonomy map:
 \beqs
 \Gamma:  \A &\rightarrow & G^{\otimes E} \nonumber\\
 A &\rightarrow &(g_{e_1},\cdots, g_{e_E}), 
 \eeqs
  where $g_e(A)\in G$ denotes the holonomy of the connection
  along the edge $e$ of the graph $\Gamma$.
which associates to any connection the holonomy of this connection
along the $E$ edges of the graph $\Gamma$. The space of
cylindrical functionals associated with $\Gamma$ is the pullback by
$\Gamma$ of  $C^\infty(G^{\otimes E})$ defined by:
\beq
\Gamma^{*}\phi(A)=\phi(g_e(A)).
\eeq
The action of the gauge group on
$\A$ translates into an  action at the vertices of the graph
$\Gamma$, if we denote by $s(e)$ and $t(e)$ the source and target
of the edge $e$ the gauge group action is given by
\beq g_e(A^k)=
k_{s(e)}^{-1}g_e(A)k_{t(e)}
\eeq
The space of gauge invariant graph connection is denoted
$A_\Gamma =G^{\otimes E } /G^{\otimes V }$ and cylindrical gauge invariant
functionals are functionals on $ A_\Gamma $.
We want to construct a measure $d\mu_\Gamma$ on $ A_\Gamma $
which provides an Hilbert space structure to the space of gauge
invariant cylindrical functional
$\H_\Gamma=L^2(\Gamma,d\mu_\Gamma)$.
This Hilbert space structure should give a representation of the
Yang-Mills operator algebra restricted to $\Gamma $. This
operator algebra denoted $O_\Gamma$ is the quantization of
the cotangent structure $T^*(G^{\otimes E}/G^{\otimes V})$
generated by the
multiplication by gauge invariant function on $G^{\otimes E }$
and by the gauge  invariant derivation operators.
We will see that this algebra is the operator algebra of a system of
gauged  particles (see section \ref{section:particle}).
The main constraint which determines (almost but not totally)
the measure $d\mu_\Gamma$ is the fact that real classical quantities
should be quantized as hermitian operators. In the case of a
compact group, there is a unique solution up to an overall factor:
the product of normalized Haar measures over each edge
group element.
\beq \label{trivmeas} d\mu_\Gamma=\prod_{e\in
E_\Gamma}dg_e.
\eeq
This measure is then consistently extended to
the space of all cylindrical functions and defines a measure on
the space of generalized connection modulo gauge invariance
\cite{al1,al4}. In the non-compact case, it is no longer possible to
integrate gauge invariant functional with \Ref{trivmeas} since the
volume of the group is infinite. In order to construct the correct
measure, we need to divide by the infinite volume of the gauge
group, hence to gauge fix the gauge group action. We will do so in
the following by showing that $A_\Gamma$ is isomorphic to
$G^{h_\Gamma-1}$ where $h_\Gamma $ is the genus -handle number-
of the surface
obtained by blowing up the graph $\Gamma$. To be precise the
isomorphism is only between dense subspaces. The measure we are
looking for is obtained as the pushforward of the Haar measure on
$G^{h_\Gamma-1}$.

\par
 The isomorphism is constructed by a gauge fixing
procedure. This gauge fixing procedure is done in two steps.
First in  section {\bf \ref{treefixing}.} we choose a maximal tree in
$\Gamma$ and we show that $A_\Gamma \sim G^{h_\Gamma}/Ad(G)$ where $Ad(G)$
denotes the adjoint diagonal action.
In section {\bf\ref{MEASURE}.} we introduce general useful facts about non compact groups
and presents some important results of algebraic geometry allowing us to
understand the geometry of non
compact quotient spaces.
Then in section {\bf\ref{Ahmeas}.}, we continue the gauge fixing by showing that there
exists an isomorphism
between (dense subsets of) $G^{h_\Gamma}/Ad(G)$
and $G^{h_\Gamma -1}$.
The isomorphism constructed being far from obvious. We finally show
that the pull back measure is independent from all
gauge choices leading to a well defined canonical measure
$d\mu_\Gamma$ on $A_\Gamma$.
In section {\bf \ref{spinnet}.} we show that Ad$G$-invariant and naively hermitic differential
operators are indeed hermitic operator for the measure $d\mu_\Gamma$, we can define
spin networks states as eigenvectors of a complete basis of such operators.
In sections {\bf  \ref{oneloop}.} and {\bf \ref{twoloops}.} we present explicit results for the rank
one groups $\su, \slc, \slr$. Section {\bf\ref{oneloop}.} is devoted to 
the case where $h_\Gamma$,
the genus of the graph, is one. This case is very different from the generic case treated in
section {\bf \ref{twoloops}.}
Finally in section {\bf \ref{Hilbert}.} we discuss the construction of  the  full Hilbert
space of all spin networks and show that it exhibits some interesting Fock substructure.

\section{Gauge Fixing Cylindrical Functions}

\label{treefixing}

\subs{Constructing Flowers}
$A_\Gamma =G^{\otimes E } /G^{\otimes V }$ is the space of graph invariant gauge
connections.
If $\Gamma=\cup_i \Gamma_i$ is a non connected graph then
$A_\Gamma$  decomposes as the cross product $ \otimes_i A_{\Gamma_i}$.
It is therefore enough to understand the construction for the case
of connected graphs and we will restrict in the following, unless
specified otherwise, to  connected graphs only.

 $\Gamma$ is composed of $E$ oriented
edges and $V$ vertices. Each oriented edge $e$ starts at the
source vertex $s(e)$ and ends at the target vertex $t(e)$.
A function on $A_\Gamma$
is a function on $G^{\otimes E}$ which satisfies gauge invariance at each vertex.
 More
precisely, given group elements $k_v$ at each vertex $v$, $\phi$
satisfies:
\begin{equation}
\phi(g_{e_i})=\phi(k^{-1}_{s(e_i)}g_{e_i}k_{t(e_i)}),\,i=1\dots E
\label{ginv}
\end{equation}
 Our first goal is to define a measure to integrate such a function.
For this purpose, we would like to identify the ``true'' degrees
of freedom of $\phi$: we are going to gauge fix the gauge
invariance \Ref{ginv}.

There is a very simple and natural gauge fixing
 for graph connections
which consists in  eliminating as many variables
$g_e$ as possible by  fixing them to, say, the identity $1$.
 More
precisely, we choose a maximal tree $T$ on our graph $\Gamma$. $T$
is a subset of edges which touches every vertex without ever
forming a loop. The characteristic property of a maximal tree is
that there exists a unique path along the tree $T$ which connects
any two given vertices of $\Gamma$. In particular, $T$ is made up
of $V-1$ edges. Given two vertices $A$ and $B$, we can define the
oriented product of group elements $h^T_{AB}$ along the path in
$T$ connecting $A$ and $B$. Now, using the gauge invariance
\Ref{ginv}, we can fix all the group elements on the edges of $T$
to 1.  To achieve this, we first need to choose a vertex $A$ from
which we are going to write our gauge fixing procedure. And we use
\Ref{ginv} with
\beq \label{fixk} k_v=h^T_{v A} \eeq
For an arbitrary edge $e$, the transformation reads
\beq
 \gt_{e}=h^T_{A s(e)} g_e  h^{T}_{t(e) A} \label{gtoG}
\eeq
 Let's consider an edge $e\in T$. There exists an unique path
in $T$ linking it to $A$, else there would be a loop in the tree
$T$. There is two situations: either the path connects $A$ to
$s(e)$ or it connects $A$ and $t(e)$. Reversing the orientation of
$e$, we can  choose for example that the path connects $A$ to
$t(e)$. Then,  $h^T_{s(e)A}=g_eh^T_{t(e)A}$ and $h^T_{A
s(e)}=(h^T_{s(e)A})^{-1}$ so that \Ref{gtoG} reads $\gt_{e}=1$.
So \ref{fixk} fixes all the group elements living on the edges of
the tree $T$ to 1.
This defines a function $\phi_T$ depending on the $g_\Gamma=E-V+1$
group elements living on the edges {\bf not} in $T$:
\beq
\phi_T( \{ \gt_e,e\notin T \} )= \phi(g_e = \gt_e \tr{ if } e\notin T
\tr{ or } = 1 \tr{ else})
\eeq
This new function has a simple residual gauge invariance:
\beq \forall k\in G, \,
\phi_T(\gt_{f_i})=\phi_T(k^{-1}\gt_{f_i}k),i=1\dots g_\Gamma
\label{flower}
\eeq
In other word, this gauge fixing procedure is an isomorphism
\beq
T:G^{h_\Gamma}/Ad(G) \rightarrow A_\Gamma
\eeq
and $\phi_T$ is the
pull back of $\phi$ by this isomorphism.

 The residual gauge invariance
corresponds to a graph with a single vertex which we call a {\it
flower}. What happens is that we have contracted the whole tree
$T$ to the single $A$, which is the remaining vertex. A priori,
this construction and therefore the function $\phi_T$ depends on
the choice of the point $A$. In fact, the whole construction is
independent of this choice. Shifting from the vertex $A$ to
another vertex $B$, we can define the product of the group
elements along the path in $T$ going from $A$ to $B$; let's note
it $h=h^T_{A B}$. The gauge fixing procedure carried from $B$ will
create variables:
\beq
\tl{G}^{(T)}_{e}=h^T_{B s(e)} g_e h^T_{t(e) B}= h^{-1} h^T_{A
s(e)} g_e h^T_{t(e)A} h= h^{-1} \gt_{e} h
\eeq
 We will define a new function $\tl{\phi}_T$ based on these new
variables, but it will be {\it equal} to $\phi_T$ due to the gauge
invariance \Ref{flower} for $k=h$.

One important issue for later  is the way the function
$\phi_T$ changes when we modify the maximal tree $T$ on which it
is based. Let's therefore choose another maximal tree $U$. We can
follow the same gauge fixing procedure based on the vertex $A$ to
define variables $\gu_e$ for each edge $e$ not belonging to $U$
and define a function on the flower $\phi_U$. To relate $\phi_T$
and $\phi_U$, we would like to decompose the variables $\gu_e$
onto the variables $\gt_e$. Let's more generally consider any
oriented loop ${\cal L}$ starting at the point $A$ and coming back
to the point $A$ and try to express the oriented product of the
group elements along it - say $H$ - in terms of the $\gt_e$. Such
a loop must contain at least an edge not belonging to $T$; else
the tree $T$ would contain a loop, which is impossible. Then, it
is easy to realize that $H$ is the oriented product -following the
orientation of the loop  ${\cal L}$- of the variables $\gt_e$ for
$e$ on ${\cal L}$ and not belonging to $T$.
For an edge $e\notin U$, the group element $\gu_e$ can be expressed as the
holonomy around the loop ${\cal L}^{(U)}[e]$
following $U$ from $A$ to $s(e)$ and back from $t(e)$ to $A$.
We can therefore
decomposes $\gu_e$ into an oriented product of $\gt_f$. Coming
back to the function $\phi_T$ and $\phi_U$, this implies that:
\beq
\phi_T(\gt_e)=
\phi_U(\gu_e=\overrightarrow{\prod_{f\in {\cal L}[e]\setminus T}} \gt_f)
\label{change}
\eeq

\subs{Examples}
In the following we give some illustration of all these procedures
for the following  simple graph:
\begin{center}
\epsfig{figure=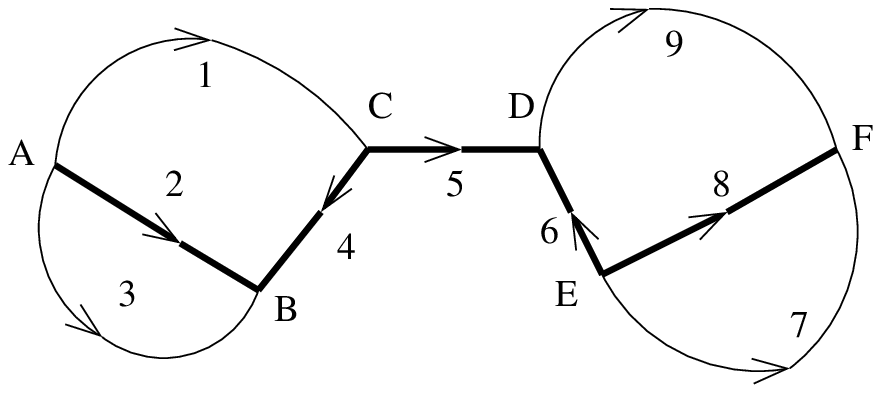,height=3.5cm}
\end{center}

\ni
The function $\phi$ satisfies the following relation
for every variables $k\in G$

\beqs
 & \phi(g_1,\dots,g_9)= \\
&\phi(k_A^{-1}g_1k_C,k_A^{-1}g_2k_B,k_A^{-1}g_3k_B,k_C^{-1}g_4k_B,
k_C^{-1}g_5k_D,k_E^{-1}g_7k_F,k_E^{-1}g_8k_F,k_D^{-1}g_9k_F)
\nonumber
 \eeqs

We choose the tree $T$ as indicated on the above graph and we write down
the gauge fixed variables $\gt$ based on the vertex $C$:

\beqs
\gt_1=g_4g_2^{-1}g_1 \quad \gt_3=g_4g_2^{-1}g_3g_4^{-1} \quad \nonumber \\
\gt_7=g_5g_6^{-1}g_7g_8^{-1}g_6g_5^{-1} \quad
\gt_9=g_5g_9g_8^{-1}g_6g_5^{-1}
\eeqs

\begin{center}
\input{flowerT.eepic}
%\hbox{\epsfig{figure=flowerT.pstex,height=3cm}}
\end{center}

\ni
Then one defines the function on the four petal flower by the
following relation:

\beqs
& \phi_T(\gt_1,\gt_3,\gt_7,\gt_9)= \nonumber\\
& \phi(\gt_1,1,\gt_3,1,1,1,
\gt_7,1, \gt_9) \\
\eeqs
One can do the same for another tree $U$:
\begin{center}
%\hbox{\epsfig{figure=treeU.eps,height=3.5cm}}
\epsfig{figure=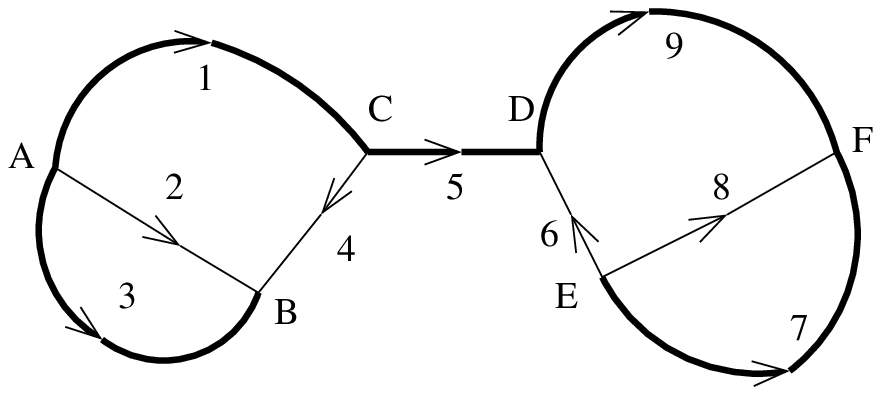,height=3.5cm}
\end{center}
Still based on the vertex $C$, we define the $\gu$ variables:
\beqs
\gu_2=g_1^{-1}g_2g_3^{-1}g_1 &\quad &
\gu_4=g_4g_3^{-1} g_1 \nonumber \\
\gu_6=g_5g_9g_7^{-1}g_6g_5^{-1} &\quad &
\gu_8=g_5g_9g_7^{-1}g_8g_9^{-1}g_5^{-1}
\eeqs
Then, we can also define the function $\phi_U$ as done above
for the tree $T$. Now, following the procedure \Ref{change} of
change of tree, we find the decomposition of the $\gu$ variables
in terms of the $\gt$ variables:

\beqs
\gu_2=(\gt_1)^{-1}(\gt_3)^{-1}\gt_1 & \quad &
\gu_4=(\gt_3)^{-1}\gt_1 \nonumber \\
\gu_8=\gt_9(\gt_7)^{-1}(\gt_9)^{-1} & \quad &
\gu_6=\gt_9(\gt_7)^{-1}
\eeqs
If one is skeptical, one
can check these relations directly using the initial $g$
variables. Finally, we get the relation between the two functions
on the flower:

\beqs
& \phi_T(\gt_1,\gt_3,\gt_7,\gt_9)= \nonumber \\
& \phi_U((\gt_1)^{-1}(\gt_3)^{-1}\gt_1,(\gt_3)^{-1}\gt_1,
\gt_9(\gt_7)^{-1},\gt_9(\gt_7)^{-1}(\gt_9)^{-1})
\eeqs

\section{Reductive groups, Quotient space and algebraic geometry}

\label{MEASURE}

We have reduced so far the problem of constructing a measure on
$A_\Gamma$ to the problem  of constructing  invariant measures on
the spaces $A_{h}= G\times \cdots \times G/ Ad(G)$, where $Ad(G)$
denotes the diagonal adjoint action  of $G$.
\beq \label{gaction}
g\cdot (g_{1},\cdots,g_{h}) \rightarrow (gg_{1}g^{-1},\cdots,
gg_{h}g^{-1}). \eeq
The measure we are seeking for should be
symmetric and invariant under right and left multiplication;
\beqs
d\mu(g_{\sigma_{1} },\cdots,g_{\sigma_{h}})  = d\mu(g_{1},\cdots,g_{h}) , \\
d\mu(kg_{1}h,\cdots,g_{h}) = d\mu(g_{1},\cdots,g_{h}), \eeqs where
$\sigma$ is a permutation. It should also satisfy reality
conditions. \\
More precisely, suppose $P(X_1,\cdots, X_h)$ is a real ($P^\dag
=P$) $Ad(G)$-invariant element of $U({\cal G}^h)$ - $U({\cal G})$ being
the universal enveloping algebra of $G$. $P$ can be realized as a
differential operator on $A_{h}$ using the correspondence between
a Lie algebra element $X$ and left invariant derivative operator:
\beq
\partial_X^i \phi(g_1,\cdot,g_i,\cdots,g_h)\equiv
\phi(g_1,\cdot,g_i X,\cdots,g_h).
 \eeq
This differential operator $P$ should be hermitian with respect to
the measure $d\mu$ we are seeking for.
In the case of a compact group, there is only one measure satisfying this conditions;
$d\mu$ is just the product of
normalized Haar measures for each group factor.
The symmetry property is obvious and the implementation of the reality conditions are
equivalent to the right and left invariance of the Haar measure.
The
integral of a gauge invariant function $\phi$ over $G^{\otimes 2}$
factorizes
\beq
\int_{G^{\otimes 2}}dg_{1}dg_{2}
\phi(g_{1},g_{2})=
 vol(G)\int_{A_{2}} d\mu(g_{1},g_{2}) \phi(g_{1},g_{2}),
\eeq
since the volume of a compact group is finite, and we can normalize the
Haar measure such that it is one.
In the case of a non compact group, this is no longer possible since
one would have to divide by the volume of the gauge group which is
obviously infinite. We therefore have to be more cautious in the
construction. \\
Before constructing explicitly the measure on $A_{h}$, we need to
introduce some results concerning Lie algebras and Lie groups (all
these facts can be studied more thoroughly in e-g \cite{Knapp}).

\subsection{Reductive Lie group and Cartan subalgebra}

The theory we are developing is valid for all, so-called, linear
connected reductive semi-simple groups i-e algebraic matrix subgroups
which are connected, invariant under conjugate transpose and of
finite center. This contains all compact Lie groups but also
non-compact ones which will be our main interest. Among these,
we distinguish complex groups ($SL(N,\C)$, $SO(N,\C)$, and
$Sp(N,\C)$) from real non compact groups (e-g $SL(2,\R)$,
$SO(N,1)$, or $SL(N,\R)$ \ldots).

Given a Lie group $G$ and its Lie algebra $\G$, a Cartan Lie
subalgebra  $\cal{H}$ is a maximal abelian subalgebra of $\G$
stable under the conjugate transpose. A Cartan subgroup $H$ is
the centralizer of a Cartan subalgebra $\cal H$ (i-e the subgroup
of element of $G$ commuting with all the elements of $\cal H$).
For each Cartan subgroup, we define the Weyl group as $W(H)
=N(H)/H$, where $N(H)$ denotes the normalizer of $H$. In the case
of compact groups, there is only one Cartan subalgebra, moreover
any group element can be conjugated to the Cartan subgroup. In the
non compact case, this is no longer true. First, in general, there
is a finite number of non conjugate Cartan subalgebras, which
all have the same rank (e-g 2 for $SL(2,\R)$,  1 for $SO(2N+1,1)$,
$N$ for $SL(N,\R)$). Note that for complex groups (e-g $SL(N,\C)$)
there is only one Cartan subgroup. Second, not all elements of $G$
can be conjugated to a Cartan subgroup. The elements which can are
called regular and the corresponding set is denoted $G_{1}$.
$G_{1} $ consists of elements $x$ such that $Ad x$ is
diagonalizable. It is an open set in $G$ and its complement is of
Haar measure $0$. Moreover the action of $Ad(G)$ on $G_{1}$ is
regular, $G_{1}/Ad(G)$ is  equal to the disjoint union
$\sqcup_{i} H_{i}/W(H_{i})$ of the
Cartan subgroups modulo their Weyl group .

Given a Cartan subalgebra $ \cal H$ we have a Cartan decomposition of
$\G$,
$\cal G= {\cal H} \oplus {\cal B(H)} $, where ${\cal H} $ is a Cartan
subalgebra and ${\cal B(H)}$ an associated Borel subalgebra.
${\cal B(H)}$  uniquely decomposes in terms of eigenspaces of
$ad{\cal H} $, i-e
${\cal B} = \oplus_{\alpha \in \Delta(H) }{\cal B}_{\alpha} $ where
$\Delta(H) $ is the root space $\Delta \subset {\cal H}^{*}$.
The root space decomposes into the union of  positive roots
$\Delta^{+}$ and negative roots $\Delta^{-}$.
The generators of ${\cal H} $ are either compact $H^{*}=-H$ or non
compact $H^{*}=H$.

Given a Cartan subgroup $H$ of $G$, we define a measure on $G/H$
as follows
\beq\label{GH}
\int_{G} f(g) dg = \int_{ G/H  }
\left[\int_{H} f(xh) dh \right]  dx,
\eeq
where $dg, dh$ are the
invariant Haar measures on $G,H$ and $f$ is a compactly supported
function on $G$. Note that $dx$ is still invariant under left
multiplication, therefore, we have the identity
\beq \label{HGH}
\int_{G} f(g) dg= \int_{H} \left[\int_{G/H} f(h^{-1}x h) dx\right]
dh.
\eeq

Note that we cannot in the RHS of \ref{HGH} innocently  interchange the
$x$ and $h$ integration as in \ref{GH}.
$G_1$ can be decomposed in an  union of conjugacy classes
$G_1= \sqcup_i {G}^H_{i}$, where
${G}^H_{i}=\{ghg^{-1},h \in H_{i},g \in G \}$, each conjugacy class
covers  $w(H_{i})=\#W(H_i)$ times a connected component of  $G_{1}$.
The integral over $G$ can therefore be reexpressed as an integral
over conjugacy classes, this is the Weyl integration formula:
\beq\label{Weyl}
\int_{G} f(g) dg = \sum_{i} \f{1}{w(H_{i})}
\int_{H_{i}} \left[\int_{G/H_{i} } f(xhx^{-1}) dx \right]
|\Delta_{i}(h)|^{2} dh,
\eeq
where
\beq
\Delta(e^{H})=  \prod_{\alpha \in \Delta^{+}(H_{i}) }
\sinh \f{\alpha(H)}{2},
\eeq
for $H \in {\cal H}_{i}^\C $, $e^{H} \in  H_{i}$.
Contrary to the case of compact groups, not all group elements
can be obtain as an exponential $e^{X}$ with $X\in \G$.
However we can realize any group element as $e^{X}$ with
$X$ in the complexified Lie algebra $ \G^\C$.

\subsection{Orbit space invariant theory and algebraic geometry}
\label{quotient}

$A_{h}$  is defined as a quotient space by the action of a group $G$.
We know that in general we do not get as a result  a nice Haussdorf
manifold.
Several types of singularity can arise when we consider orbit spaces.
Let's look for instance at the case of $A_{2}=G\times G/Ad(G)$.
If $(g_{1},g_{2})$ are generic (non-commuting) elements of $G$,
the isotropy group of these points is the center of $G$ hence a finite
subgroup.
But, if $g_{1}$ and $g_{2}$ are commuting elements of $G$,
the isotropy group is non trivial.
It is the intersection of the centralizer of $g_{1}$ and
$g_{2}$.
If say $g_{1}$ is regular, it is a Cartan subgroup
and in general the dimension of the isotropy subgroup is, at least,
the rank of the group.
These non-generic points can act like attractors for the action of $Ad(G)$.
Suppose, for instance, that $G=SL(2,\R)$ and
$(g_{1},g_{2})= (1,e^{\sigma_3})$  where $\sigma_{3} = diag(+1,-1)$.
 The  isotropy group of this point is
 the group $ e^{t\sigma_3}$. This point is non Haussdorf and is an attractor
for some neighboring orbits. To see that, lets  consider
$g_u= (e^{u\sigma_{+}}, e^{\sigma_{3}}) $
where  $ \sigma_{+}=\mat{0}{1}{0}{0} $. Then
$\lim_{t\rightarrow \infty} e^{-t\sigma_{3}} g_u e^{t\sigma_{3}}
=(1,e^{\sigma_3})$. The orbit $Ad(G).{(e^{u\sigma_{+}},
e^{\sigma_{3}})}$ is not closed since it contains the fixed point
$(1,g_{2})$. Therefore any neighborhood of the orbit $Ad(G).{(e^{u\sigma_{+}},
e^{\sigma_{3}})}$ contains $(1,e^{\sigma_{3}})$.
So two different orbits associated with $u>0$ and $u^\prime<0$ have non
disjoint neighborhoods which means that the quotient space is not
Haussdorf.
One way to cure this problem is to exclude from the beginning the set
of commuting elements, so that all orbits are closed.
However this is not enough.
Suppose, for instance,  that $G=SL(2,\R)$ and denote by $(x,y) \in
\R^{2}$ the group element
$(e^{\sigma_{3}}e^{x\sigma_{+}},e^{\sigma_{3}} e^{y\sigma_{-}})$.
The action of $e^{t\sigma_{3}}$ translates into the action
$(x,y)\rightarrow (e^t x,e^{-t}y)$. The exclusion of the commuting
elements translates into the condition $(x,y)\neq (0,0)$.
In this space, all orbits are closed. But one can see that any
neighborhood of the orbit of $(x,0)$ will
intersect a neighborhood of the orbit of $(0,y)$ and the quotient is
once again not Haussdorf.
The solution is to exclude the points $(x,0)$ or $(0,y)$, then we
obtain a nice quotient.

This example illustrates the general problematic in defining quotient
spaces. In fact, since $G$ is an algebraic group (being a subgroup of matrices)
and since the $Ad(G)$ action is also algebraic, this problem has received
a lot of attention in the mathematical literature when the group is
complex \cite{Brion,Spring} under the name of invariant theory.

\medskip

First, one needs to recall general facts from algebraic geometry and
then give the definition of a regular or {\it geometric quotient space}.
An affine algebraic variety $X$ over $\C $ is defined as being the set of zeros of a
collection of polynomials of $\C^{N}$, $X= \cap_{i}V(P_{i})$, where $V(P)=\{x \in
\C^N \ P(x) =0 \}$.
The topology which is useful in this context is the Zariski topology
where the  closed sets are the algebraic subvariety of $X$, generated
by $X(P)=X\cap V(P)$. The open sets are finite union of standard open
sets $X_{P}= \{ x\in X, P(x)\neq 0\}$. It is important to note that
the open sets in Zariski topology are much bigger than in the usual
topology. For instance any open set of $X$  is a dense subset of $X$
and any finite intersection of open sets is also dense.
Given an affine algebraic variety $X$, one defines the algebra of
{\it regular} functions, denoted $\C[X]$, as the algebra of polynomials on
$\C^N$ restricted to $X$, it is clear that $\C[X]= \C[\C^N]/I(X)$,
where $I(X)$ is the ideal of polynomials which are zero on $X$. \\
One of the basic theorem in this context is that any subalgebra $A$ of
$\C[\C^N]$(or any commutative algebra) which is finitely generated and
does not
contain nilpotent elements is the algebra of regular functions on an
affine variety $X$. Such an algebra is called affine.
$X$ is called the spectrum of $A$ and defined as the set of
homomorphism $A\rightarrow \C$.
This theorem translates geometry into algebra. \\
Given an affine variety $X$, we say that it is irreducible if it
cannot be decomposed as the union of two subvarieties. In the algebraic
language, this translates into the condition that $\C[X]$ is integral.
Given $\C[X]$, we can define the field of rational functions and we
denote it $\C(X)$.
We say that an application between two affine varieties $\phi:X \rightarrow
Y$  is a morphism iff $\phi^{*}$ maps regular functions of $Y$ onto regular
functions of $X$.

\medskip

We are now ready to precise the good notion of a quotient. Roughly a
good quotient space is one for which orbits are separated by rational
invariant functions.
Let $G$ be an algebraic group acting on an affine irreducible variety  $X$.
A geometrical quotient of $X$ by the action of $G$ is an affine variety
$Y$ together with a surjective morphism $\pi: X\rightarrow Y$ such
that:
\begin{itemize}
\item (i) $\pi$ induces an isomorphism between  $\C(Y)$ and
$\C(X)^{G}$.
\item (ii) The fibers of $\pi$ are the orbits of $G$ in $X$.
\end{itemize}

The condition (ii) tells us that $Y$ is a quotient space since
it is a orbit space. The condition (i) tells us that this quotient space is a
 algebraic variety where points are separated by rational functions.
We saw in the previous examples that
such a good quotient space does not exist in general.
Hopefully there is a
fundamental {\bf theorem of Rosenlich} \cite{Brion}
which states that given a variety $X$ and an algebraic action of $G$ on
$X$, it is always possible to choose an open dense set $X_{0}$ stable
under $G$ such that $X_{0}/G$ is a good quotient.
The proof of this theorem goes as follows. We first restrict $Y$ such that
hypothesis (i) is satisfied. Then, hypothesis (i) implies
that the orbit of $x$ is dense in $\pi^{-1}(x)$.
However, in general, hypothesis (ii)
is not true. What we do next is to restrict ourself to a subset $X_0$ of $X$
which contains only orbits of maximal dimension. This implies (ii).
Then the geometric quotient $X_{0} /G$ exists as an algebraic variety.

\medskip

In the case the group $G$ is reductive, there exists a fundamental
theorem due to Hilbert and Nagata which states that if $X$ is
irreducible and $G$ reductive then $\C[X]^{G}$ is finitely generated.
Since $\C[X]^{G}$  does not contain nilpotent elements, this means that
it is an affine algebra and therefore that it is the algebra of
regular functions over its spectrum, which is denoted
$X//G \equiv spec(\C[X]^{G})$. It is equipped with a surjective morphism
 $\pi:X\rightarrow X//G $ and called the {\it quotient} of $X$ by
$G$.
The quotient of $X$ by $G$ is universal in the sense that  any
$G$-invariant morphism $p:X\rightarrow Y$ can be factorized over
$X//G$, i-e there exists a morphism $q: X//G\rightarrow Y$ such that
$p= q\circ \pi$.
It is then possible to show
that any fiber of $\pi$ contains a {\it unique} closed orbit.
Geometrically this means that $X//G$ is the space of closed orbits
of $G$ in $X$.
This is a little bit disappointing since this means that $X//G $ could
be a very rough description of the space of orbits.
For instance $X//G$ is generally not a geometric quotient.
For instance Let $X=\C^{2}$ and $G=\C^{*}$ acts by multiplication
$(x,y)\rightarrow (tx,ty)$.
The only invariant polynomials $P(x,y)$ are the constant polynomials,
so that $X//G$ is reduced to a point. In other words, there is one
unique closed orbit, the one of $(0,0)$.
Fortunately, the following property is true if $G$ is linear reductive
connected and {\it semisimple}. In this case, the algebra of fractions
of $\C[X]^{G}$ (i-e $\C(X//G)$) is equal to $\C(X)^{G} $.
Equivalently this means that $X//G$ is the space of dense orbits, i-e
any  fiber  $\pi:X\rightarrow X//G $ contains a dense orbit.
Therefore this theorem imply that in the case we consider, e-g linear
reductive group we can define the geometrical quotient space $X/G$ as the
algebraic dual of the space of invariant polynomial.

\medskip

Coming back to our specific problem
we have $G$ a linear reductive  connected semi-simple Lie group acting
on $X=G^{n}$ by the adjoint action.
By the general theory just exposed, we know that,
in the case $G$ is complex, the universal orbit space
$G^{n}//Ad G$ consisting of dense orbits separated by invariant
polynomials on $G^{n}$ is well defined. Moreover by the
Rosenlich theorem, we know that it is always possible to exclude from
$G^{n}$ a closed set such that the universal orbit space is well defined as a
geometric quotient.
However the drawbacks of these general methods, despite their beauty
and generality, are that they say nothing about the real case in which
we are interested and that they are not constructive. \\
It is therefore still interesting to understand better and explicitly  which
closed set we have to exclude from $G^{n}$ in order to get  a geometric quotient
space (which is then a well defined affine variety). Moreover we are
interested in the measure theoretical property of the quotient and we
would like to show that the difference between the geometrical
quotient and the quotient space $  G^{n}//G$ is of measure zero. \\
First we know from Rosenlich theorem that it is possible to obtain a
geometrical quotient by excluding  a closed (in the Zariski topology)
set from $G^n$. \\
In the case of $G/Ad(G)$, this problem is well known. The solution is
to exclude from $G$ the points at which the $Ad(G)$ action is not regular.
The set of regular elements of $G$ is denoted by $G^\prime$ and this is
the set of $g\in G$ for which $Ad(G)$ is diagonalizable. The quotient
space  $G^\prime/Ad(G) $ is  equal to the union of the
Cartan subgroups modulo their Weyl group $\sqcup_{i} H_{i}/W(H_{i})$
(the intersection of $H_{i}$ with $H_{j}$ only contains the identity).
We need to introduce the set of strictly regular
elements of $G$ denoted $G_{1}$ and which is the set of regular elements
of $G$ not equal to the identity, $G^\prime = G_{1}\cup \{Id\}$.

In the case of $G\times G$, the same strategy is working and we need to
exclude from it non regular points.
We therefore consider the action  of $Ad(G)$ not on $G\times G$ but on
a subspace. For a general group even if we know the existence of such a
maximal subspace we don't have an explicit  description and this deserve
a full mathematical study. We can however give an explicit description in
the case of rank 1 groups, in that case we define
\beq \label{g2def}
G_{2}\equiv \{ (g_{1},g_{2}) \in G \times G ; g_{1}\in G_{1}\,
\textrm{ or }
\,g_{2}\in G_{1} \textrm{, and }
g_{1}g_{2}g_{1}^{-1}g_{2}^{-1}\in G_{1} \},
\eeq
We then have the following proposition
\begin{prop}
\label{G2}
$G_{2}$ is a dense subset of $G\times G$,
its complement is of Haar
measure zero and $G_{2}/Ad(G)$
is a geometric quotient when $G$ is of rank one.
Therefore, $A_{2}=G_{2}/Ad(G)$ is an Haussdorf manifold of dimension
$dimG$ which separates rational functions
and it is  the  base manifold  of  a homogeneous fiber bundle
whose fiber is $G$ and total space $G_{2}\approx A_{2}\times G $.
\end{prop}
This proposal is proved in the section \ref{twoloops} where we
construct the dual spectrum space
to the space of invariant polynomials and show that
it is isomorphic to $G_2/Ad(G)$. The main point is that the condition of
 being in $G_1$ can be implemented as an algebraic inequality.

Note that, in the definition \ref{g2def}, $G_2$ is such that the
centralizer of any element of $G_2$ is trivial. Suppose $g$
commutes with $(g_1,g_2)\in G_2$, since we can take e-g $g_1$ to
be regular hence $g$ can be diagonalized in the same basis as $g_1$
(the regularity assumption is essential here). $g$ cannot be
regular since it commutes with $g_2$ and therefore it would mean
that $g_2$ is also diagonal in the same basis, hence commutes with
$g_1$, which is impossible. So the definition of $G_2$ implies that
$g$ is diagonal and non-regular. If the rank of the group is 1,
this means that $g$ is the identity. Whereas if the rank is higher, we
clearly need additional conditions to get the same conclusion. It
is important to understand that the definition of $G_2$ is not the
naive definition where one just excludes from $G\times G$ the group
elements which have a non trivial centralizer: in order to get a
nice quotient, we need to take away more points and this is
dictated by the fact that we want the quotient space to be the
spectrum of the algebra of invariant functionals.

In the general case of higher rank group we define $G_2$ to be the following space
\beq \label{2G2}
G_2\equiv\{(g_{1},g_{2}) \in G \times G ; g_{1}\in G_{1}\,
\textrm{ or }
\,g_{2}\in G_{1} \textrm{, and }
C(g_{1},g_{2}) = Z_G \},
\eeq
where, $C(g_{1},g_{2})$ denotes the centralizer of $g_1$ and $g_2$ and $Z_G$
denotes the center of the group.
We will see in the next section that $G_2$ admits a quotient by $Ad(G)$ and we denote by
$A_2$ the quotient.
However, contrary to the definition \ref{g2def} valid for rank 1 group,
the definition \ref{2G2} is not equivalent to a definition of $G_2$ as an algebraic
dual.

\section{construction of the measure on $A_h$}
\label{Ahmeas}

This section is central to our paper.
In this section we go back to our problem which is the construction of a
measure on $A_h$. We show, by a non trivial gauge fixing procedure,
that a dense subset of $G^h/Ad(G)$ can be identify with a dense subset of $G^{h-1}$
when
$h > 1$. we first consider $A_2$, the case of one Cartan subgroup, then the case of
several Cartan subgroups, we then consider the case of $A_h$.
Finally, putting everything together we show that the measure on $A_\Gamma$ does
not depend on all the gauge fixing choices, leading to the definition of a canonical
measure.

\subsection{Construction of the measure on $A_2$}

\sss{Case of a unique Cartan subgroup $H$}
\label{gaugefixingproc}
Let's consider the following embedding of $G_{1}$ into $A_{2}$.
Given $g \in G_{1}$, we can conjugate it to the Cartan subgroup $H$
of $G$ i-e there exists $h\in H$, $x \in G/H$ such that
$g=xhx^{-1}$.
We choose a  section $s:G/H \rightarrow G$ and
we define the map:

\begin{equation}
\begin{array}{cccc}
j_{s}: &G_{1} &\rightarrow& A_{2} \\
& g=xhx^{-1} & \rightarrow & Ad(G).(h,s(x))
\end{array}
\end{equation}
where $Ad(G).(h,s(x))$ is the orbit of $(h,s(x))$ under the
conjugation by $G$.
This is a gauge fixing since given $(g_1,g_2)\in A_2 $
we can conjugate $g_1$
to the Cartan subgroup $H$ of $G$
i-e there exists $h\in H$, $y \in G/H$ such that
$g_1=yhy^{-1}$. Then $Ad(G).(g_1,g_2)=Ad(G).(h,y^{-1}g_2y)$.
This fixes only partially the gauge since
$H$ can act on $y$ by $y\rightarrow yk$ which means
that we can still conjugate $\tl{g}_2=y^{-1}g_2y$ by a Cartan group
element.
Nevertheless, since we are in $A_2$, the centralizer of
$g_1$ and $g_2$ is trivial, this means that the centralizer of
$h\in H$ and $\tl{g}_2$ is trivial.
Since the centralizer of $h \in H\cap G_1$ is $H$,
this implies that the conjugate action of $H$ on
$\tl{g}_2$ has no other fixed point than the center elements.
Let's suppose for the following that the center of $G$ is trivial.
Then, given an arbitrary section $s:G/H \rightarrow G$,
we can use the residual symmetry coming from the conjugation by $H$
to impose that $\tl{g}_2$ belongs to the image of $s$. Finally,
the gauge fixing we impose is
$(g_1,g_2)\rightarrow(h,s(x)) \in H\times G/H $.
We just have argued that every element of $G_2$ can be brought to this form.
Moreover the condition that the centralizer of $(g_1,g_2)$  is implemented if
we ask $g=xhx^{-1} \notin H$.
Indeed, $xhx^{-1} \in H$ would mean that either
$x\in H$ or that $x$ is a Weyl transformation. The first possibility,
$x\in H$ is impossible since $x$ and $h$ don't commute.
So that $j_s$ gives a map from  $G_1\setminus H$ to $A_2$.
The second possibility
is related to the Gribov ambiguity, which makes
the definition of $j_s$ still ambiguous.

This can be traced back to the fact that
a given group element $g$ can be conjugated to
different Cartan elements related by the action of the Weyl
group $W(H)$, which is the residual conjugation action on
the Cartan subgroup.
There is two ways to solve this problem.
First, we can require good
transformations of the section $s$ under the action of the Weyl
group

\beq
\forall x\in G/H,\,\forall w\in W(H),\,s(xw)=w^{-1}s(x)w
\eeq
This renders $j_{s}$ well-defined and this is the hypothesis
we will suppose in the following.
Or we can impose $h$ to be in a fixed Weyl chamber. In this case,
one must remove all the $1/w(H)$ factors from the following proofs.

With this map, we can pullback functions on $A_{2}$, or
equivalently invariant functions on $G_2$, to functions on
$G_{1}$ by $j_{s}^{*}F (xhx^{-1}) = F(h,s(x))$.

\begin{defi} \label{defmeasure}
Let $\mu$ be a measure on $A_{2}$ defined by
\beq
\int_{A_{2}} F(g_{1},g_{2}) d\mu(g_{1},g_{2})\equiv
\int_{G} j_{s}^{*}F (g) dg.
\eeq
\end{defi}

\begin{prop} \label{facto}
Let $F$ be a $L^1$ function on $G_2$ with respect to the Haar measure.
We also require that its gauge invariant version $^GF$ is well-defined:
\beq
^{G}F(g_{1},g_{2})=\int_{G} F(gg_{1} g^{-1}, gg_{2} g^{-1})  dg
\eeq
Then we have
\beq
\int_{G\times G} F(g_{1},g_{2}) dg_{1} dg_{2}
=\int_{A_{2}} {}^GF(g_{1},g_{2}) d\mu(g_{1},g_{2}),
\eeq
\end{prop}

Let $F(g_{1},g_{2}) $ be a $L^1$ function on $G\times G$.
\beqs
& &\int_{G\times G} F(g_{1},g_{2}) dg_{1}dg_{2}\\
&=&\f{1}{w(H)}\int_{G/H\times H\times G}
F(xhx^{-1}, g_{2}) \Delta(h) dx dh dg_{2}\\
&=& \f{1}{w(H)}\int_{G/H\times H\times G}
F(xhx^{-1},x g_{2}x^{-1}) \Delta(h) dx dh dg_{2}
\eeqs
where we  have used the Weyl integration formula (\ref{Weyl}) in the
first equality and the invariance under right and left translation of
the Haar measure $dg_2$ in the second.
Using the identity  (\ref{HGH}) for the integration on $G_{2}$ the
integral can be expanded as
\beqs
& & \f{1}{w(H)}
\int_{G/H\times H\times H\times H\backslash G  }
F(xhx^{-1}, xkyk^{-1}x^{-1}) \Delta(h) dx dh dk dy \nonumber\\
&=& \f{1}{w(H)} \int_{ H\times H\backslash G }
\left[\int_{G/H\times H} F(xhx^{-1}, xky(xk)^{-1})  dx dk  \right]
\Delta(h) dh dy \nonumber \\
&=& \f{1}{w(H)} \int_{ H\times H\backslash G }
\left[\int_{G/H\times H} F(xkh(xk)^{-1}, xky(xk)^{-1})  dx dk  \right]
\Delta(h) dh dy
\eeqs
where we have used the fact that $H$ is abelian
to derive the last equality.
Then using the definition of the
$G/H$ measure (\ref{GH}), we finally get
\beq
\int_{G\times G} F(g_{1},g_{2}) dg_{1}dg_{2} =
\f{1}{w(H)} \int_{ H\times H\backslash G }
 {}^{G}F(h, y)
\Delta(h) dh dy
\eeq
where $^{G}F$ is the gauge invariant version of $F$:
\beq
^{G}F(g_{1},g_{2})=\int_{G} F(gg_{1} g^{-1}, gg_{2} g^{-1})  dg.
\eeq

\begin{theo}\label{theorem}
$\mu$ is  independent of $s$, symmetric,invariant under
right and left multiplication and invariant under taking the inverse:
\beqs
d\mu(g_{1},g_{2}) = d\mu(g_{2},g_{1}) \\
d\mu(kg_{1}h,g_{2}) = d\mu(g_{1},g_{2}) \\
d\mu(g_{1},g_{2}) = d\mu(g_{1}^{-1},g_{2})
\eeqs
\end{theo}

Let's prove the above theorem for $d\mu(kg_{1},g_{2})$ (left
multiplication). One will be able to prove the other properties
following the same line of thoughts.
The easiest way to prove the theorem
is to use a Faddev-Popov gauge fixing procedure
using the proposition \ref{facto}.
Let's choose an invariant
function $F$ on $G_2$.
We can choose any function
$\varphi$ on $G_2$ such that $^G\varphi=1$ and
create the (gauge fixed) function
$\tilde{F}=F\varphi$.
In the usual Faddeev-Popov procedure one would choose
$\varphi$ to be proportional to a $\delta$
function of a gauge fixing condition,
but this is not necessary.
Applying proposition \ref{facto}:

\beq
\int_{A_{2}} F(g_{1},g_{2}) d\mu(g_{1},g_{2})
=\int_{G\times G} \tilde{F}(g_{1},g_{2}) dg_{1} dg_{2},
\eeq

Using the freedom in the choice of the function
$\varphi$ in proposition \ref{facto}
and the fact that
if $\varphi(g_1,g_2)$ is a gauge fixing so
is $\varphi_k(g_1,g_2)=\varphi(k^{-1}g_1,g_2)$,
we get
\begin{eqnarray}
\int_{A_{2}}  F(g_1,g_2)\dmu(kg_1,g_2)
&= & \int_{G\times G}  F(k^{-1}g_1,g_2)
\varphi(k^{-1}g_1,g_2) dg_1 dg_2 \nonumber \\
&=& \int_{G\times G}  F(g_1,g_2)\varphi(g_1,g_2)dg_1 dg_2 \nonumber \\
&=& \int_{A_{2}} F(g_1,g_2)\dmu(g_1,g_2)
\end{eqnarray}
So we conclude to the left invariance of the measure $\dmu$
defined on $A_2$.

\sss{The case of many Cartan subgroups}

In general, we have many Cartan subgroups and let's note them
$H_1,H_2,\dots,H_n$. $G_1$ can be decomposed in disconnected
components $G^{(i)}=Ad(G).H_i$,
each conjugated to the Cartan subgroup $H_i$.
For each of these components, one can choose a section
$s_i:G/H_i\rightarrow G$ and define a map

\begin{equation}
\begin{array}{cccc}
j_i: &G^{(i)} \subset G_1 &\rightarrow& A_{2} \\
& g=yhy^{-1} & \rightarrow & Ad(G).(h,s_i(y))
\end{array}
\end{equation}
From this map, one can define a measure $d\mu_i$ on $A_2$
as in the definition \Ref{defmeasure} by

\beq
\int_{A_{2}} F(g_{1},g_{2}) d\mu_i(g_{1},g_{2})\equiv
\int_{G_i} j_i^{*}F (g) dg
\eeq

\begin{prop}
Given any $L^1$ function $F$ on $G_{2}$  we
have
\beq
\int_{G\times G} F(g_{1},g_{2}) dg_{1} dg_{2}
=\int_{A_{2}} {}^GF(g_{1},g_{2}) d\mu(g_{1},g_{2}),
\eeq
where
\beq
d\mu(g_{1},g_{2})=\sum_i \f{1}{w(H_{i})}d\mu_i(g_{1},g_{2})
\eeq
and where $^{G}F$ is the gauge invariant version of $F$:
\beq
^{G}F(g_{1},g_{2})=\int_{G} F(gg_{1} g^{-1}, gg_{2} g^{-1})  dg
\eeq
\end{prop}

Let $F(g_{1},g_{2}) $ be a $L^1$ function on $G\times G$.
\beqs
& &\int_{G\times G} F(g_{1},g_{2}) dg_{1}dg_{2}\\
&=&\sum_{i}\f{1}{w(H_{i})}\int_{G/H_{i}\times H_{i}\times G}
F(xhx^{-1}, g_{2}) \Delta(h) dx dh dg_{2}\\
&=& \sum_{i}\f{1}{w(H_{i})}\int_{G/H_{i}\times H_{i}\times G}
F(xhx^{-1},x g_{2}x^{-1}) \Delta(h) dx dh dg_{2}
\eeqs
where we  have used the Weyl integration formula (\ref{Weyl}) in the
first equality and the invariance under right and left translation of
the Haar measure in the second.
Using the identity  (\ref{HGH}) for the integration on $G_{2}$ the
integral can be expanded as
\beqs
& & \sum_{i}\f{1}{w(H_{i})}
\int_{G/H_{i}\times H_{i}\times H_{i}\times H_{i}\backslash G  }
F(xhx^{-1}, xkyk^{-1}x^{-1}) \Delta(h) dx dh dk dy\\
&=& \sum_{i}\f{1}{w(H_{i})} \int_{ H_{i}\times H_{i}\backslash G }
\left[\int_{G/H_{i}\times H_{i}} F(xhx^{-1}, xky(xk)^{-1})  dx dk  \right]
\Delta(h) dh dy
\eeqs
Using the fact that $H_{i}$ is abelian and the definition of the
$G/H_i$ measure (\ref{GH}) we finally get
\beqs
\int_{G\times G} F(g_{1},g_{2}) dg_{1}dg_{2} &= &
\sum_{i}\f{1}{w(H_{i})} \int_{ H_{i}\times H_{i}\backslash G }
 {}^{G}F(h, y)
\Delta(h) dh dy \nonumber \\
&=&\sum_{i}\f{1}{w(H_{i})} \int_{G^{(i)}}
dg \,\, j_i^* ({}^GF)(g) \nonumber \\
&=&\sum_{i}\f{1}{w(H_{i})} \int_{A_2} d\mu_i(g_1,g_2) \,
{}^GF(g_1,g_2)
\eeqs

Using the above proposition, we can generalize theorem
\ref{theorem} to the multi-Cartan case using the same proof as
done before.

\subs{The measure on $A_h$}

We are now interested in generalizing the case of $A_2$ to
$A_h$. Applying the Rosenlich theorem stated in \Ref{quotient},
it is always possible to choose an open dense set $G_h\subset G^h$
such that the geometric quotient $A_h=G_h/Ad(G)$ is well-defined
as in proposition \ref{G2}. Following the choice made for the
2-petals case we get the following definition when $G$ is rank
one.
\begin{defi}
\beq
G_{h} \equiv \{ (g_{1},\cdots ,g_{h}) \in G^{\otimes h}|\,
  \exists
(i,j) \in [1,\cdots,h],\, (g_{i},g_{j}) \in G_{2} \}
\eeq
\end{defi}
Note that $G_h$ is such that the centralizer of any element in
$G_h$ is the identity.

\begin{defi} \label{defdmun}
We choose two edges $i,j$ on the $n$-petal flower. Then, we can define
a measure on $A_h$
\beq
\mu^{(ij)}\left[f(g_1,g_2,\dots,g_n)\right]
=\int_{G_2^{(ij)}} d\mu(g_i,g_j)
\int\prod_{k\ne i,j}dg_k
f(g_1,g_2,\dots,g_n)
\eeq
where we have taken the measure gauge fixed measure $d\mu(g_i,g_j)$ on
the two chosen edges and the Haar measure on the other edges.
\end{defi}
This measure is well-defined since $\int\prod_{k\ne i,j}dg_k
f(g_1,g_2,\dots,g_n)$ is an invariant function on $G_2^{(ij)}$
(and therefore a function on $A_2$, which we can integrate
using $d\mu(g_i,g_j)$).

\begin{prop} \label{facton}
Given any $L^1$ function $F$ on $G^n$  we
have
\beq
\int_{G^n} F(g_{1},\dots,g_n) dg_{1} \dots dg_n
=\int_{A_{n}} {}^GF(g_{1},\dots,g_n) d\mu(g_{1},\dots,g_n),
\eeq
where $^{G}F$ is the gauge invariant version of $F$:
\beq
^{G}F(g_{1},\dots,g_n)=\int_{G} F(gg_1 g^{-1}, \dots,gg_n g^{-1})  dg
\eeq
\end{prop}
This proposition is easily proved using the proposition \ref{facto}
and the invariance
of the Haar measure under right and left multiplication. And its leads
to the following theorem:

\begin{theo} \label{totalinv}
The measures $d\mu^{(ij)}$ don't depend on the choice of edges $i,j$.
And one defines a unique measure $d\mu(g_1,g_2,\dots,g_n)$ on $A_h$.
Moreover, this measure is symmetric under permutation of $g_1,\dots,g_n$,
under right and left multiplication and under taking the inverse of one
of its argument:

\beqs
d\mu(g_{\sigma_{1} },\cdots,g_{\sigma_{n}})  = d\mu(g_{1},\cdots,g_{n}) ,
\nonumber \\
d\mu(kg_{1}h,\cdots,g_{n}) = d\mu(g_{1},\cdots,g_{n}),
\nonumber \\
d\mu(g_{1}^{-1},\cdots,g_{n})=d\mu(g_{1},\cdots,g_{n})
\eeqs
\end{theo}

\subs{Measure on an arbitrary graph}

To construct the measure on an arbitrary graph $\Gamma$,
we are going to choose
a maximal tree $T$ and carry on the gauge fixing procedure
described in the first section
in order to reduce the graph $\Gamma$ to a flower.
We then define the measure $\dmu_T$ such
that for all gauge invariant functions $\phi$ on $\Gamma$, we have:

\beq
\int \dmu_T(g_1,\dots,g_E) \phi(g_1,\dots,g_E) =
\int \wtl{\dmu}(g_1,\dots,g_F) \phi_T(g_1,\dots,g_F)
\eeq
where $\wtl{\dmu}$ is the measure on the $F$ petal flower.

\medskip

This definition a priori depends
on the choice of the tree $T$. We are going to prove
that this is not the case. So we choose
two maximal trees $T$ and $U$. The gauge fixed
function are related by \Ref{change}:

$$
\phi_T(\gt_e)=
\phi_U(\gu_e=\overrightarrow{\prod_{f\in {\cal L}[e]\setminus T}} \gt_f)
$$
and we want to prove that:

\beq
\int \wtl{\dmu}(\gt_1,\dots,\gt_F) \phi_T(\gt_1,\dots,\gt_F)
=
\int \wtl{\dmu}(\gu_1,\dots,\gu_F) \phi_U(\gu_1,\dots,\gu_F)
\eeq
or equivalently:

\beq
\int \wtl{\dmu}(\gu_1,\dots,\gu_F) \phi_U(\gu_1,\dots,\gu_F)=
\int \wtl{\dmu}(\gt_1,\dots,\gt_F)
\phi_U(\gu_e=\overrightarrow{\prod_{f\in {\cal L}[e]\setminus T}} \gt_f)
\label{chvar}
\eeq
We are going to show this equality by doing some
elementary changes of variables
which would correspond to elementary moves between the
two maximal trees and we will show that the measure
is invariant under each such move.
Let's first define what we mean by an elementary move.

\begin{defi}
Given a graph $\Gamma$ and a maximal tree $T$ on it,
let's choose a vertex $v$  such that there is at least one edge linked to it
which is not in the tree $T$. Let's pick one and call it $f$.
There exists an unique path in $T$ linking the other vertex of $f$ to $v$.
This path goes along an unique edge $e\in T$ linked to $v$.
Then an elementary change of tree, or elementary move, is exchanging the role
of $e$ and $f$ and considering the maximal tree $U=T\cup f \setminus e$.
\end{defi}

\begin{center}
\input{move.eepic}
\end{center}

\ni
The interest in such a definition
lies into the following proposition~:

\begin{prop}\label{elmoves}
Having chosen two maximal trees $T$ and $U$ on a graph $\Gamma$, there exists
a sequence of elementary moves going from $T$ to $U$.
\end{prop}

Then, as we will see,
the change of variable from $G^{(U)}$ to $G^{(T)}$ is very
simple for such a move since it is implemented either by an inversion
or by left multiplication, so that it will simplify the study
of change of trees.

\medskip

Let's first prove the proposition. Given $\Gamma$ and two maximal
trees $T,U$ in it we can distinguish four types of edges:
edges belonging to both trees $T$ and $U$, edges in $V=T\setminus
U$, edges in $W=U\setminus T$ and edges in neither trees. By
elementary moves on either the tree $T$ or the tree $U$, we would
like to reduce the sets $V$ and $W$ down to nothing. Let's take a
closer look at the set $V$. First, $V$ might not be connected. In
this case, we would carry on the following procedure on each of
its connected parts.  Let's denote one of the connected part $V_1$
and work on it. $V_1$ is a tree as part of the tree $T$. In
particular, it is not closed and has some open ends i.e edges
connected to $V_1$ only by one vertex. By doing elementary moves,
we are going to remove them from $V_1$, and then, by repeating the
same operations, one could erase completely the set $V_1$. And
finally, by repeating the procedure on the other connected parts
of $V$, one could absorb completely the set $T\setminus U$.

So let's choose an edge $e$ at an open end of $V_1$. It has two vertices:
$v$ in the exterior of $V_1$ and $w$ in the interior of $V_1$.
There exists an unique path ${\cal P}$ along
$U$ which links these two vertices and
$e \notin U$
is not in it.
In ${\cal P}$, there exists at least one edge in $U$
but not in $T$ else there would
be a loop in the tree $T$.

Let's suppose that such an edge $f\in U\setminus T$
touches directly the edge $e$
(at the vertex $v$).
Then, we can do an elementary move exchanging $e\lrarr f$ and create a tree
$\tl{U}=U\cup e \setminus f$ closer to the tree $T$ than the tree $U$.

\begin{center}
\input{move2.eepic}
\end{center}

\ni
Let's now come back to the general case in which we have to
follow a sequence of edges $f_1, \dots,f_n \in T\cap U$ starting
from the vertex $v$ along the path ${\cal P}$ to an edge $f$ in
$U\setminus T$. Then, we do the allowed elementary moves on the
tree $U$ exchanging $e\lrarr f_1, \dots, f_{n-1}\lrarr f_n$, thus
creating the trees $U_1,\dots U_n$. Starting from $v$, all the
edges $e,f_1,\dots, f_{n-1}$ are both in $T$ and $U_n$, $f_n$ is
in $T\setminus U_n$, $f$ is in $U_n\setminus T$, and all the other
edges on the way back to $w$ are in $U_n$. So that we are in the
same simple case as above and we finally do the move $f_n\lrarr f$
on the tree $U_n$ creating a tree $\tl{U}$ such that the whole
loop ${\cal P}$ from $v$ to $v$ is in both $T$ and $\tl{U}$ except
the edge $f$ which is in neither. Practically, we had a loop with
all the edges in $U$ but one which is in $T$ (it's $e$), and by
elementary moves, we move it around until it meets an edge which
is not in $T$ and they ``cancel'' each other.

This ends the absorbtion of the edge $e$:
the set $T\setminus \tl{U}$ contains
one edge less than $T\setminus U$. And we now repeat
the same procedure using the
new tree $\tl{U}$.

\medskip
We now are able to prove that~:

\begin{theo}
\label{moveind}
The Jacobian of the change of variables \Ref{chvar} is 1, so that the measure
$\dmu_T$ is invariant under changes of tree.
\end{theo}

This theorem will assure the existence
of a {\bf measure $\dmu^{(\Gamma)}=\dmu_T$
independent from the choice of the tree $T$ and therefore from the whole
gauge fixing procedure}, which is the measure
we will use to integrate our gauge
invariant functions and define a space
of $L^2$ gauge invariant functions. This space
will in fact be the Hilbert space of spin networks
defined on the graph $\Gamma$.

\medskip

Proposition \ref{elmoves} means that we only have to prove the
theorem \ref{moveind} for elementary moves.
So, let's realize an elementary move
on the tree $T$ around the vertex $v$ and define the new maximal
tree $U=T\cup f \setminus e$. For every edge $a\notin U$ on the
$U$-flower, we define the variables $\gu_a$. And we want to
express them in terms of the variables $\gu_b$. For $a\notin U$ and
$\notin T$, we want to relate $\gu_a$ to $\gt_a$. It can be easily
seen that  these two variables are equal up to a multiplication on
the left or right or on both side by $\gt_f$ or its inverse. And
for the only other case $a=e$, we will have $\gu_e=(\gt_f)^{\pm
1}$. Then using the invariance of the measure of the flower by
multiplication or by taking the inverse of one of its argument
(theorem \ref{totalinv}), we can conclude that the above change of
variables has a trivial Jacobian.

\section{Spin Networks States}
\label{spinnet}

In this section, we are going to define the spin networks as eigenvectors
of a set of commuting Laplacian operators, which will be shown to be hermitian.

\subs{Laplacian operators}

Let's consider a graph $\Gamma$ and a gauge invariant function
defined on it.  These functions depend on $E$ group elements
$g_1,\dots,g_E$.
Let's denote by $X$ an element of the Lie algebra and by
$\pp_{X}^{R_e}$ (resp. $\pp_{X}^{L_e}$)
the right (resp. left) invariant derivative acting on the $j$-th
group element associated with the edge $e$~:

\beqs
\pp_{X}^{R_1}f(g_{1},\cdots,g_{N}) =f(Xg_{1},\cdots,g_{N})\\
\pp_{X}^{L_1}f(g_{1},\cdots,g_{N}) =f(g_{1}(-X),\cdots,g_{N})
\eeqs
The gauge group action acts on derivative
operators by conjugation at each vertex~:

$$
\pp_{X}^{R_e} \rightarrow \pp_{k_{s(e)}Xk_{s(e)}^{-1}}^{R_e}.
$$

We are interested in gauge invariant differential operators.
The algebra of such operators is generated by
Laplacian operators $\Delta^{(i)}_e$ where
$e$ labels the edges of $\Gamma$ and $i$ runs from 1 to
the rank $r$ of the group.
For each edge $e$, the space of Laplacians is in one to one correspondence
with the Casimir operators of the Lie algebra.
Therefore this set of Laplacians gives
a complete basis of commuting operators.
Indeed, they are commuting since for a given $e$
two Casimirs commute and for different edges $e$'s the
differential operators $\pp_{X}^{R_e}$ commute with each other.

We want to define the spin networks as the basis
of eigenstates vectors for this complete set
of gauge invariant differential operators.
In order to do that, we need to show that these operators
are hermitian with respect to the
measure $d\mu^{(\Gamma)}$ that we have just constructed.
We are going to give the proof for the quadratic Laplacian operator
$\Delta_e=\sum_i\pp_{X_i}^{R_e}\pp_{X_i}^{R_e}$
($\Delta_e = \pp^{R_e}.\pp^{R_e}$ for short),
where $X_i$ denotes an orthonormal basis of the Lie algebra.
The general case is similar, it simply needs more cumbersome notations.

\medskip

Because of the measure $\dmu^{(\Gamma)}$ has being defined on the
gauge fixed flower corresponding to $\Gamma$,
we have to follow the gauge fixing procedure leading to group variables
$G_i,\dots,G_F$ on the flower and express the operators $\Delta_e$ in term
of the derivatives $\tl{\pp}^{L,R}_i$ with respect to these new variables.

To start with, let's look at an example:
the case of the two petal flower coming from
either the $\Theta$ graph or the eyeglass graph.
Let's first gauge fix the $\Theta$ graph:

\ni
\begin{minipage}{8cm}
\begin{center}
\input{theta.eepic}
\end{center}
\end{minipage}
\begin{minipage}{5cm}
\input{2petal.eepic}
\end{minipage}

\ni
We gauge fix from the point $A$. We have
$G_1=g_1g_3^{-1}$ and $G_2=g_2g_3^{-1}$.
It is then easy to check that
$\pp^R_1=\tl{\pp}^R_1$, $\pp^R_2=\tl{\pp}^R_2$
and $\pp^R_3=\tl{\pp}^{L_1}+\tl{\pp}^{L_2}$
so that $\Delta_1=\tl{L}_1$, $\Delta_2=\tl{L}_2$
and $\Delta_3=\tl{\Delta}_1+\tl{\Delta}_2+2\tl{\Delta}_{12}$,
where $\tl{\Delta}_{12}=\tl{\pp}^{L}_1.\tl{\pp}^{L}_2$.

In the case of the eyeglass graph:

\begin{center}
\input{eyeglass.eepic}
\end{center}
We have $G_1=g_1$ and $G_2=g_3g_2g_3^{-1}$. Thus
$\Delta_1=\tl{\Delta}_1$, $\Delta_2=\tl{\Delta}_2$
as the Laplacian is invariant under
$Ad(G)$ and $\Delta_3=(\tl{\pp}_2^R-\tl{\pp}_2^L)^2$.

In the generic case, for a given edge $e$,
if there exists a maximal tree $T$ which doesn't go
through $e$, then $e$ will be on the gauge fixed flower
and $\Delta_e$ will simply be the Laplacian
$\tl{\Delta}_e$ with derivatives with respect to
the flower variable $G_e$.

What happens to edges which are
in every possible trees, such as the middle edge in the eyeglass graph,
is slightly more tricky. For such an edge $e$, we choose
to gauge fix from its departure vertex $v$. Then $\pp_e^R$ will be
equal to the sum of $\tl{\pp}_f^R$ for all edges $f$
whose loop starts with $e$
and $\tl{\pp}_f^L$ for all  edges  whose loop finishes with $e$.

\medskip
In any case, the initial differential operators
$\Delta_e$ can be written as a sum
of $\tl{\Delta}_{ij}^{RR}=\tl{\pp}^{R}_i.\tl{\pp}^{R}_j$ oand
$\tl{\Delta}_{ij}^{LR}=\tl{\pp}^{L}_i.\tl{\pp}^{R}_j$
where $i,j =1,\cdots,g$. These operators are $Ad(G)$ invariant
operators on $G^h$.
This is easily seen since the gauge invariance of
a function $\phi_\Gamma$ is equivalent to

\beq
(\sum_{e|s(e)=v} \pp^{R_e} + \sum_{e|t(e)=v} \pp^{L_e}) \phi=0,
\eeq
for all vertices $v$.
Choosing a tree amounts to use these equations
to express all derivatives along edges belonging to the tree in terms of
the other ones that we named $\tl{\pp}_i^{L,R}$.
We are then left with only one relation
$\sum_{i=1}^h (\tl{\pp}_i^{R}+ \tl{\pp}^{L}_i)\phi =0 $

\subs{Spin Networks as Laplacian eigenvectors}

\begin{theo}
The Laplacian operators $\tl{\Delta}_{ij}^{RR}$,
$\tl{\Delta}_{ij}^{LR}$ are hermitian
with respect to the measure
$\dmu_h$, $h>1$.
\end{theo}
We will give the proof for
$\tl{\Delta}_{i}\equiv\tl{\Delta}_{ii}^{RR}$. The proof for a
general operator is similar.

Let's consider
\beq
\label{sym}
\int_{A_{h}} (\varphi\tl{\Delta}_i \psi -\psi\tl{\Delta}_i \varphi) d\mu_h
\eeq
where $\varphi,\psi$ are gauge invariant functions.
And let's introduce a gauge fixing function $\phi$, which is such that
\beq\label{gfix}
\int_{G} \,{}^g\phi dg =1
\eeq
where
${}^g\phi(g_{1},\cdots, g_{N})
= \phi(gg_{1}g^{-1},\cdots, gg_{N}g^{-1})$.
The integral (\ref{sym}) can be written as
\beq
\int_{G^{h}}(\varphi\tl{\Delta}_i \psi -\psi\tl{\Delta}_i \varphi)
 \phi \,dg_{1}\cdots dg_{h}
\eeq
Using the invariance of the Haar measure under left multiplication we
can integrate by part the right invariant derivatives, this leads to
(we take off the $\tl{}$ for simplicity of the notations)~:

\beq
\int_{A_{h}} (\varphi\Delta \psi -\psi\Delta \varphi) d\mu =
\int_{G^{h}} \psi\pp_{X_{j}}^{R_i}\varphi \pp_{X_{j}}^{R_i}\phi
-\varphi\pp_{X_{j}}^{R_i}\psi \pp_{X_{j}}^{R_i}\phi \,dg_{1}\cdots dg_{h}
\eeq
Let's look at the first term, we can write it as
\beq
\int_{G^{h}} dg_{1}\cdots dg_{h} \,
\psi\pp_{X_{j}}^{R_i}\varphi \pp_{X_{j}}^{R_i}\phi
=
\int_{A_{h}} \psi
\left[
\int_{G}dg\, {}^g\pp_{X_{i}}^{R_k}\varphi {}^g\pp_{X_{i}}^{R_l}\phi
\right]d\mu_h
\eeq
where we have used the definition of the invariant measure \Ref{facton}
and $^g\psi=\psi$.
Using the following identity
\beq
{}^g\pp_{X_{i}}\phi = \pp_{Ad(g)^{-1}\cdot X_{i}} {}^g\phi,
\eeq
and the invariance of the quadratic differential operator
\beq
 \sum_{i} \pp_{Ad(g)\cdot X_{i}}^{R} \otimes \pp_{Ad(g)\cdot X_{i}}^{R}
 = \sum_{i} \pp_{X_{i}}^{R}\otimes \pp_{X_{i}}^{R}
\eeq
one gets
\beq
\int_{A_{h}}
\psi
\left[
\int_{G}dg  \pp_{X_{i}}{}^g\varphi
\pp_{X_{i}}{}^g\phi \,dg
\right]\dmu_h.
\eeq
Using that ${}^g\varphi=\varphi$ since $\varphi$ is
gauge invariant and that the
condition (\ref{gfix}) implies
$\pp_{X_{i}}\int {}^g\phi dg =0$, we conclude that
the integral (\ref{sym}) is zero.

\begin{defi}
Since the operators $\Delta_e$ form a set of commuting hermitian operators
on the Hilbert space $L^2(\dmu^{(\Gamma)})$,
we can diagonalize them and their eigenvectors
form an orthonormal basis of $L^2(\dmu^{(\Gamma)})$.
We call these eigenvectors {\bf spin networks}.
\end{defi}

It should be clear that these vectors should be considered as
generalized $\delta$ normalizable vectors if one of the eigenvalues
they carry is part of a continuous spectrum~: one
should consider them as invariant distributions and not invariant
functions.

We obtain a basis of functions
which are labelled by the eigenvalues of the Laplacians - the
Casimirs of the group - on each edge of the graph
$\Gamma$. In other words, if we call $d\rho(\lambda)$ the
spectral measure of the Laplacian operator one gets

\beq
H_\Gamma
\equiv L^2(\dmu^{(\Gamma)})= \oplus_e \int
d\rho(\lambda_e)\otimes_v I_v(\lambda),
\eeq
where $I_v(\lambda)$
is the space of intertwiners between the
representations carried by the edges meeting at
 the vertex $v$.

In the case of $G=SU(2)$, there is a one-to-one correspondence
between the eigenvalues of the Casimir and the irreducible unitary
representations. And by the previous reasoning, we have totally
reconstructed the usual structure of spin networks with
representations labelling the edges of the spin networks if the
graph is trivalent. In a more general context, one should be
careful that the eigenvalues of the Laplacians do not always
completely characterize the representations, there can be a
degeneracy where several representations carry the same Laplacians.
This is the case, for instance, of the series of discrete
representations of $\slr$, where the degeneracy is $2$. This is not
the case however for the unitary representations of $\slc$ which
are totally determined by the values of their two Casimirs.
We have presented the
spin networks in the particular case of rank $1$ but as we said in
the beginning, all the propositions work the same for a more
general group.

\subs{Unfolding vertices and $SU(2)$ spin networks}

Using the gauge fixing procedure, we can unfold all the vertices
of a given graph $\Gamma$.
By this, we mean replace each vertex by a (minimal)
tree which has only 3-valent vertices. More explicitly,
let's consider a vertex $v$. We can match the edges meeting at $v$
two by two (if their number is odd, we leave one edge on its own) and
create a 3-valent vertex for each of these pairs. Then we repeat
this process until over.

Let's call the unfolded graph $\Gamma_0$. As the flowers
corresponding to $\Gamma$ and to $\Gamma_0$ are the same, we have
$L^2(\dmu^{(\Gamma)})=L^2(\dmu^{(\Gamma_0)})$. We can then
construct spin networks on the graph $\Gamma_0$, by labelling all
its edges - both the ones already in $\Gamma$ and the new ones
which are {\it inside} the vertices of $\Gamma$ - with
representations of the group $G$. These spin networks span
$L^2(\dmu^{(\Gamma_0)})$ and therefore also
$L^2(\dmu^{(\Gamma)})$. That way, we have unfolded the structure
of  the vertices of the spin networks based on the graph $\Gamma$.
We have reduced the problem of characterizing nodes to the case of
3-valent ones. One thing which one should be aware of is that for
high rank groups, it happens that the space of trivalent
interwiners is infinite dimensional e.g $sl(3,\R)$.

In the case of $G=SU(2)$, the 3-valent nodes are
3-valent intertwiners
intertwining between the representations labelling the three edges.
These intertwiners are unique to a normalization.
So we have fully characterized $SU(2)$ spin
networks - both their edges and nodes- by the above
unfolding procedure. In the general case, we can have many
possible 3-valent intertwiners and their space has to be studied
in order to fully characterize the nodes of the spin networks.

\section{The one loop case}
\label{oneloop}

In this section, we restrict ourself to rank one groups and
 we deal with the graph made of a single loop
with a single bivalent vertex, which describes the quotient
space $G/Ad(G)$. This case is essentially different from the
cases of flowers with higher $h\ge2$ number of petals for
which the quotient space $G^h/Ad(G)$ can be mapped onto
$G^{h-1}$ as described in the previous section.
And both the techniques used and the problems
encountered differ from the oter cases.
Nevertheless, this study is important since the characters
of the unitary representations are supposed to be
orthonormal vectors of $L^2(G/Ad(G))$, therefore
it is interesting as the simplest part of the gauge
invariant connection space and
it illustrates the problem of the
possible non-connectivity of the
quotient space and the (super)selection rule issue which
comes with it \cite{Gomb}.

\subs{$SL(2,\C)$}

In this section, we consider the case where the group is $SL(2,\C)$
and we are interested in describing the quotient
$(SL(2,\C))//Ad(SL(2,\C))$ as defined in section (\ref{quotient}).
Let $g=\mat{a}{b}{c}{d} \in \slc$, the  algebra $\C[\slc]^{\slc}$
of polynomials invariant under the adjoint action is generated by
$X(g)=(1/2)tr(g)$, since such polynomials are linear combination
of $tr(g^n)$. $tr(g^n)$ can be expressed as a polynomial in $X$
due to the relation $g^{2}-tr(g) g +1=0$, more precisely $tr(g^n)
=T_{n}(X)$ with $T_{n}$ the Chebichev's polynomials of the first
kind. Therefore the spectrum of the invariant polynomial affine
algebra is just $\C$ and the quotient  morphism is the trace.
Moreover  $tr^{-1}(x), x \neq \pm 1$ is exactly an orbit of a
strictly regular element and $tr^{-1}(\pm 1)= \{\pm Id\} \cup
\{\pm\mat{1}{z}{0}{1}| z\in\C\} \cup \{\pm\mat{1}{0}{z}{1}|
z\in\C\}$. One can therefore think of $\slc//Ad(\slc)$ as the
geometric quotient of $G_{1}\cup \pm Id$. The G-invariant measure
(Weyl measure) induced by the $\slc$ Haar measure is given by

\beq \label{dhslc}
\mu(f)=\int_{\C}|X^{2}-1|f(X) dX
\eeq
where the integration region
is over the complex plane minus the interval $[-1,+1]$
with the usual Lebesgue measure on $\C$.

\medskip

More explicitly, $\slc$ has only one Cartan subgroup $H$ which is
the set of diagonal matrices $diag(\lambda,\lambda^{-1})$,
$\lambda\in \C$.
The Weyl group is $Z_2$ and $diag(\lambda,\lambda^{-1})$ is
conjugate to $diag(\lambda^{-1},\lambda)$.
The Weyl integration formula reads:

\beq
 \int_{\slc} f(g) dg =
 \int_{H} dh\,\left[\int_{\slc/H} f(xhx^{-1}) dx \right]
 |\Delta(h)|^{2}
\eeq
The invariant measure is obtained by removing the redundant integration
over $\slc/H$ and integration solely on $H$ and one finds back the measure
\Ref{dhslc}.

\medskip

The unitary principal series of $\slc$ is a family of unitary
irreducible representations of $\slc$ indexed
by pairs  $(j,\rho)$ with $j\in \Z/2$ and $\rho \in \R$.
There are realized in $L^{2}(\C)$ and the action
$R_{j,\rho}$ of $\slc$ is given by
\beq
R_{j,\rho}\mat{a}{b}{c}{d} f(z) = |bz+d|^{-2-2i\rho}\left(\f{bz+d}
{|bz+d|}\right)^{2j} f\left(\f{az+c}{bz+d}\right)
\eeq
for $z\in \C$ and $f\in L^{2}(\C)$.
The characters are
\beq
\chi_{j,\rho}\mat{e^{x+i\theta}}{0}{0}{e^{-x-i\theta}} =
\f{e^{i\rho x}e^{i j\theta}+e^{-i\rho x}e^{-i j\theta}}{|e^{x+i\theta} - e^{-x-i\theta}|^2}.
\eeq
Using the measure \Ref{dhslc} on $X=(e^{ x + i \theta}+e^{-x-i
\theta})/2$ and making a change of variables to $x,\theta$, it is
straightforward to check that \beq
\mu\left(\chi_{j_1,\rho_1}\chi_{j_2,\rho_2}\right)=
\delta_{j_1j_2}\delta(\rho_1-\rho_2) \eeq so that the above
characters form a orthonormal basis of
the Hilbert space
$L^2(A_1)=L^2(\slc//Ad(\slc))$.

\subs{$SU(2)$}

In the case of $SU(2)$, there is again a unique Cartan subgroup $H$,
composed of the diagonal matrices
$h_\theta=diag(e^{i\theta},e^{-i\theta}), \theta\in[-\pi,\pi]$.
The Weyl group is $Z_2$: $h_\theta$ and $h_{-\theta}$ are conjugate.
The $SU(2)$-invariant measure is
\beq
\mu_{SU(2)}(f)=\f{2}{\pi}\int_{-1}^{+1}dX \sqrt{1-X^{2}}f(X)
=\f{2}{\pi}\int_{0}^{\pi}d\theta\sin^2\theta f(\theta)
\eeq
where $X=1/2tr(g)=\cos\theta$.

An orthonormal basis of $L^2(SU(2)/Ad(SU(2)))$ is given by the
characters of the irreducible (finite-dimensional) representation
of $SU(2)$: \beq \chi_{j}(h_\theta) =
\f{\sin(j+1)\theta}{\sin(\theta) } \eeq where j runs over the
non-negative integers (twice the spin). For making a change of
variable from $X$ to $\theta$, it is easy to check that

\beq
\mu_{SU(2)}(\chi_j\chi_k)=\delta_{jk}
\eeq

\subs{$SL(2,\R)$}

In the case of $SL(2,\R)$, we have two Cartans subgroups; a compact
one, which corresponds to space rotations
\beq
H_0=\left\{k_\theta
= \mat{\cos\theta}{\sin\theta}{-\sin\theta}{\cos\theta}, \,
0\le\theta\le 2\pi\right\}
\eeq
and a non-compact one, which
corresponds to boosts
\beq
H_1=\pm\left\{a_t=\mat{e^t}{0}{0}{e^{-t}},\,t\in R\right\}.
\eeq
$W(H_0)$ is trivial but $W(H_1)=Z_2$ and $a_t$ is conjugate to
$a_{-t}$. A regular element of $SL(2,\R)$ can be conjugated to
$H_0$ or to $H_1$ and a $Ad(SL(2,\R))$ invariant function $f$
will be described by its action on both Cartan subgroups i.e by
two functions $f_0(\theta)$ and $f^\pm_1(t),t\ge0$.

\medskip

We would like to divide by the volume of $G/H_0$ and of
$G/H_1$. These two volumes are infinite and the ratio of these
two volumes is also infinite.
So this leads to an ambiguity and we are left with a
one-parameter family (up to normalization)
of possible $Ad(SL(2,\R))$ invariant measures
\beq
\label{dmusu}
\mu_{SL(2,\R)}(f)=\alpha_0\int_0^{2\pi}d\theta\sin^2\theta f_0(\theta)
+\alpha_1\int_{0}^{+\infty}dt\sh^2t f^\pm_1(t)
\eeq
The formal property that we want our measure to satisfy is:
\beq
\label{1loopinv}
\mu(^Gf)=\int_G dg \,f(g)
\eeq
where $^Gf$ denotes the averaging over the gauge group $G$
of a compact supported non invariant
function $f$.
The subtle point is that the centralizer of a
generic group element under the adjoint
action is (conjugated to) either $H_0$ or $H_1$
depending on the group element.
So the averaging should take this into account.
Therefore if $f_0$ (resp $f_1$)is supported on the
space $G_{(0)}$ (resp $G_{(1)}$) of group elements
which can be conjugated to $H_0$ (resp $H_1$), we define
\beq
^Gf_i(g=xh_ix^{-1})=\int_{G/H_i}dx_i\,f(x_ih_ix_i^{-1})
\eeq
In this case, we easily prove using \Ref{Weyl} that
\Ref{1loopinv} is satisfied for the unique choice
$\alpha_i=1$.

\medskip

However, this is not the whole story.
The natural way to get
an invariant measure is to choose
a cutoff $\lambda$ and $G_\lambda$ a compact subset of $G$,
with $G_\lambda \rightarrow G$ when $\lambda$ grows to
infinity.
Then we take the invariant measure to be the limit
\beq
\mu(f)=\lim_{\lambda \rightarrow \infty}
\f{\int_{G_\lambda }f(g) dg}
{\int_{G_\lambda} dg},
\eeq
for a $G$ invariant function $f$. The resulting measure
that it leads to is $(\alpha_0=1, \alpha_1=0)$. And this measure
gives a zero weight to function with support on $G_{(1)}$.

The way to reconcile these points of view is the
following. One needs to define
two Hilbert spaces, one (denoted ${\cal H}_0$) for the
functions with support on $G_{(0)}$ which is given by the measure
$(\alpha_0 =1, \alpha_1=0)$ and one (denoted ${\cal H}_1$)
for the functions with support on $G_{(1)}$ which is
given by the measure $(\alpha_0 =0$, $\alpha_1=1)$.
This would take into account the fact that
the space of $Ad(G)$ invariants is disconnected with incommensurable
volume of centralizer. Physically, this means that the two sectors
cannot communicate i-e we can not find physical operators
mapping physical states between the two sectors.
This was
rigorously shown by Gomberoff and Marolf \cite{Gomb} in a similar
context but in the language of group averaging and rigging maps.

\medskip

$\slr$ has three series of principal unitary
representations: the continuous series ${\cal C}_s$ labelled by a
positive real number $s$ and two discrete series ${\cal D}^\pm_n$
both labelled by a integer $n\ge1$ and a sign.
The characters of the continuous series are

\beq
\chi_s(k_\theta) =  0
\eeq
\beq
\chi_s(\pm a_t) = \f{\cos s t}{|\sh t|}
\eeq
and the characters of the discrete series ${\cal D}^\pm_n$ are
\beq
\chi^\pm_n(k_\theta) = \mp \f{e^{\pm i(n-1)\theta}}{2i\sin\theta}
\eeq
\beq
\chi^\pm_n(a_t) = \f{e^{-(n-1)|t|}}{2|\sh t|}
\textrm{ with a factor $(-1)^n$ for } -a_t
\eeq
It is clear that the characters of the discrete series (resp.
continuous series) are orthonormal with respect to the Hilbert space
structure ${\cal H}_0$ (resp. ${\cal H}_1$). Moreover both
characters are eigenvalues of the Laplacian. More explicitly,
the Laplacian reads~:

\beq
\Delta = \f{1}{\sin\theta}\f{\pp^2}{\pp\theta^2}\sin\theta +\f{1}{4}
\textrm{ on } H_0 \qquad
\Delta = - \f{1}{\sh t}\f{\pp^2}{\pp t^2}\sh t +\f{1}{4}
\textrm{ on } H_1 \textrm{ for }t\ge 0
\eeq
so that the eigenvalue of $\chi_s$ is $s^2+1/4$ and the
one corresponding to $\chi^{\pm}_n$ is $m(1-m)$ with
$m=n-1/2$. Moreover, one could notice that, for a generic measure
\Ref{dmusu} with arbitrary $(\alpha_0,\alpha_1)$, the discrete
characters $\chi_n^\pm$ are not orthonormal, which
would be in contradiction with the fact that the Laplacian
is Hermitian, unless we restrict ourself to the choice
${\cal H}_0$ i.e $(\alpha_0=1,\alpha_1=0)$
which appear to be the only self-consistent choice
of measure when taking into account the discrete series.

\medskip

The characters of the continuous series are fully characterized as
distributions which are
eigenvectors of the Laplacian, invariant under the Weyl group
(residual gauge symmetry) and with full support only on $G_{(1)}$.
However, this is not the case for the discrete series. There are
several distributions which are both eigenvector of the Laplacian and
invariant with support in $G_{(0)}$. The solution of
this puzzle lies in the definition of the Laplacian and
more particularly in the space of functions on which
it is defined.
Indeed, among all the invariant distributions,
one should choose the ones which are not only a
distribution on $G_{(0)} \coprod G_{(1)}$
but on the full group $G$: one
asks $\chi^\pm_n$ to be an eigenvalue of the Laplacian as a
distribution on $G$. More precisely,
in order to satisfy the eigenvalue equation
$\int_G \chi(g) \Delta f(g) dg = \lambda \int_G \chi(g) f(g) dg$,
one needs to integrate by parts. If $f$ is a compact
supported function on $G_{(0)} \coprod G_{(1)}$, all the boundary terms
vanishes trivially. However, if $f$ is a compact supported function on
$G$, the vanishing of the boundary terms leads to some boundary
conditions on $\chi$. The eigen-distributions that can be extended
to distributions on $G$ are called {\bf regular}. Now it is easy to
check that such distributions (normalizable with respect to the
scalar product on ${\cal H}_0$) are in one-to-one correspondence
with unitary representations. This was first shown by
Harish-Chandra and was the foundation of his works on harmonic
analysis on non-compact group \cite{Vara}.

To sum up, the issue is about the domain of definition
of the Laplacian $\Delta$. We make it act on the Hilbert spaces
${\cal H}_0$ and ${\cal H}_1$ for consistency of the group
averaging, but the eigenvalue problems is well-defined for
distributions on the whole group $G$ (taking into account
the null elements, which are not regular). Nevertheless, we can
conclude that the discrete characters $\chi_{n}$ restricted
to $G_{(0)}$ form a basis of ${\cal H}_0$ and the
continuous characters $\chi_s$ form a basis of ${\cal H}_1$.
And the tail of $\chi_n$ on $G_{(1)}$ is due to non-trivial
boundary conditions in the eigenvalue problem.

The case of one petal graph is quite complicated, this is
essentially due to the fact that the space $A_1$ is not connected,
since we have excluded all null rotations and that taking them into
account is not straightforward.
Fortunately, as  we shall now see,  the situation for higher loop graphs is simpler
since  the generic centralizer of a point of
$G^h$ is $G$ for all elements.

\section{The two petal flower: Examples}
\label{twoloops}

\subsection{$SL(2,\C)$}

In this section, we consider the case where the group is $SL(2,\C)$
and we are interested in describing the quotient $(SL(2,\C))^{2}
//Ad(SL(2,\C))$ as defined in section (\ref{quotient}). Let
$(g_{1},g_{2}) \in \slc^{2}$ and  denote $X_{1}(g_{1},g_{2})=
(1/2) tr(g_{1})$, $X_{2}(g_{1},g_{2})= (1/2) tr(g_{2})$ and
$X_{3}(g_{1},g_{2}^{-1})= (1/2) tr(g_{1}g_{2})$. This defines an
$Ad(\slc)$-invariant morphism $\pi: \slc^{2}\rightarrow \C^{3}$.
We have the following property
\begin{prop}
   $\pi$ gives an isomorphism between the algebra of invariant
   polynomials $\C[\slc^{2}]^{\slc}$ and $\C[X_{1},X_{2},X_{3}]$.
\end{prop}
{\bf Proof:}
Let $$G_{2}(\slc)=\{(g_{1},g_{2})\in \slc | tr(g_{1})^{2}\neq 4
\textrm{ or }
tr(g_{2})^{2}\neq 4,{\mathrm and}\,tr([g_{1},g_{2}]_G) \neq 2 \}.$$
The image of
this set by $\pi$ is the complement in $\C^{3}$ of
$\Delta$, where  $\Delta$ is the closed  subset of $\C^{3}$ such that
the polynomials $X_{1}^{2} -1$ or $X_{2}^{2}- 1$, and
$\Theta(X_{1},X_{2},X_{3})\equiv (X_{3}-X_{1}X_{2})^{2}
-(X_{1}^{2}-1)(X_{2}^{2} -1)$ are equal to zero
($[,]_G$ denotes the group commutator).
This is clear since $ tr([g_{1},g_{2}]_G) -2 = 4\Theta(X_{1},X_{2},X_{3})$.
The key point is that this gives an isomorphism between
$G_{2}(\slc)/\slc$ and $ \C^{3}\setminus\Delta$.
We can construct explicitly the inverse map, Let
$s(\vec{X})=(s_{1},s_{2})$ be defined by:
\beqs
 s_{1}(\vec{X}) &=& \mat{ X_{1}+\sqrt{X_{1}^{2}-1} }{0}
 {0}{ X_{1}-\sqrt{X_{1}^{2}-1} }, \\
 s_{2}(\vec{X}) &=&
\mat{X_{2} -\f{ X_{1}X_{2}-X_3}{\sqrt{X_{1}^{2}-1}}}
 {1}
 {- \f{\Theta(\vec{X})}{X_{1}^{2}-1} }
 {X_{2} + \f{ X_{1}X_{2}-X_3}{\sqrt{X_{1}^{2}-1}} }.
\eeqs
One should be careful since $s$ needs the definition of a square
root, this means that this is a multivalued function on
$\C^{3}\setminus\Delta$,
however any change in the choice of the square root
determination is implemented by a gauge transformation, therefore $s$
is well defined as a function valued into $G_{2}(\slc)/\slc$.
Suppose  we define $\tilde{s}(\vec{X})$ the map corresponding to
another determination of the square root, i-e the one obtained from
$s$ by replacing $\sqrt{X_{1}^{2}-1}\rightarrow -\sqrt{X_{1}^{2}-1}$
we have that
 \beq
 \tilde{s}(\vec{X}) = Ad\mat{0}{i}{i}{0}\cdot s(\vec{X}).
 \eeq
It is easy to see that $\pi \circ s$ is the identity mapping on
$\C^{3}\setminus\Delta$.
It is also true that $s\circ \pi$ is the identity mapping
on $G_{2}(\slc)/\slc$.
First, given $(g_{1},g_{2})\in  G_{2}(\slc)$ we can diagonalize
$g_{1}$ since it is regular. This does not fix completely the action
of the gauge group since one can still act by a diagonal gauge
transformation and a Weyl transformation (i-e $g_{1}\rightarrow
g_{1}^{-1}$). Any diagonal transformation $diag(\lambda,\lambda^{-1})$ is acting on
$g_{2}=\mat{a}{b}{c}{d}\rightarrow \mat{a}{\lambda^{2}b}{\lambda^{-2}c}{d}$. Now,
$tr(g_{1}g_{2}g_1^{-1}g_2^{-1}) -2 = -(\lambda -\lambda^{-1})^2 bc$. The condition
$\Theta \neq 0$ translates into $bc\neq 0$, so that
one can fix the residual action by asking $b=1$.
\begin{prop}
     The invariant  measure $\mu$ defined in definition (\ref{defmeasure})
     is simply the Lebesgue measure in $\C^{3}$ when translated in
     terms of $X_{1},X_{2},X_{3}$. More precisely, let $F$ be a
     function on $\C^{3}$, $\pi^{*}F$ is an invariant function and
     \beq
     \int_{(\slc)^{2}//\slc}\pi^{*}F(g_{1},g_{2}) d\mu(g_{1},g_{2})
     =\int_{\C^{3}} F(\vec{X}) d^{2}X_{1}d^{2}X_{2}d^{2}X_{3}.
     \eeq
\end{prop}
{\bf Proof:}
Let us recall that the Haar measure  for $\slc$ is defined  as
$dg = d^{2}a d^{2}b d^{2}c d^{2}d$ $\delta^{2}(ad-bc-1)$ if
$g=\mat{a}{b}{c}{d}$.
Let
\beq
y= \mat{a}{1}{c}{d},\,
h=\mat{\lambda}{0}{0}{\lambda^{-1}},\,
g=yhy^{-1}.
\eeq
The measure on $A_{2}$ is defined by $d\mu =dg$
it is easy to see that
\beq
dg = |\lambda-\lambda^{-1}|^{2} d^{2}(\lambda+\lambda^{-1}) d^{2}a 
d^{2}d.
\eeq
Moreover $X_{1}= \lambda+\lambda^{-1}$, $X_{2}=a+d$ and
$X_{3}=\lambda a +\lambda^{-1}d$ thus
$dg =  d^{2}X_{1}d^{2}X_{2}d^{2}X_{3}$.

\medskip

Let's define the invariant functionals
\beq
\Phi_{\vec{j},\vec{\rho}}(g_{1},g_{2}) =
\chi_{j_{1},\rho_{1}}(g_{1})
 \chi_{j_{2},\rho_{2}}(g_{2})
 \chi_{j_{3},\rho_{3}}(g_{1}g_{2})
\eeq
An explicit computation gives
\beq
\int \Phi_{\vec{j},\vec{\rho}}  d\mu(g_{1},g_{2})
= \prod_{i=1}^{3} \delta(\rho_{i}) \delta_{j_{i}}
\eeq

\subsection{$SU(2)$}

We can deduce the $SU(2)$ case from the previous
formalism. We have the constraint
$\vec{X}(g_{1},g_{2}) \in I_{3}$ where
\beq
I_{3} \equiv \{ X_{i} \in ]-1,1[, \,
\Theta(\vec{X})=(X_{3}-X_{1}X_{2})^{2}
-(X_{1}^{2}-1)(X_{2}^{2} -1)<0 \}
\eeq
and the invariant measure is
\beq
\label{su2mes}
\int_{I_3}d\vec{X}.
\eeq

Writing $X=\cos\theta$, we can re-express the above constraint in term of
$\theta_{1,2,3}\in[0,\pi]$:
\beq
\cos(\theta_1+\theta_2)\le\cos\theta_3\le\cos(\theta_1-\theta_2)
\eeq
which is simply the constraint arising when multiplying two
elements of $SU(2)$ or  equivalently summing two vectors in a
spherical space. Indeed two group elements $g_1,g_2$ in
$SU(2) \sim S^3$
determine a triangle in $S^3$ with vertices $1,g_1,g_2$. The
invariant geometry of this triangle is determined by the three
edges length which are  $\theta_1, \theta_2, \theta_3$.
In these variables, the measure is
$\sin\theta_1 \sin\theta_2 \sin\theta_3 d\theta_1d\theta_2d\theta_3$
and the domain of integration is
\beqs
\theta_1 +\theta_2 \leq \theta_3\,\textrm{and cyclic perm.} \\
\theta_1 +\theta_2 + \theta_3\leq 2\pi.
\eeqs
One can also express the invariant geometry of the triangle in
terms of two edges: their lengths $\theta_1, \theta_2$ and the angle
$\tl\theta_3$ they form. This angle is determined by the edges
lengths by $\cos\theta_3=cos\theta_1\cos\theta_2
-\sin\theta_1\sin\theta_2 \cos\tl\theta_3$. In these geometric
variables, the condition on the variables reads
$\Theta=-\sin^2\theta_1\sin^2\theta_2\sin^2\tl\theta_3 \neq 0$
which means that we exclude degenerate triangles. In these
new variables,
the measure is $\sin^2\theta_1 \sin^2\theta_2 d\theta_1d\theta_2
\sin\tl\theta_3 d\tl\theta_3$. Now one can easily check that \beq
\int_{I_{3}} \chi_{j_{1}}(X_{1})
\chi_{j_{2}}(X_{2})\chi_{j_{3}}(X_3) d\vec{X}=
\delta_{j_{1},j_{2}} \delta_{j_{2},j_{3}} \f{1}{d_{j_{3}}}.
\eeq
as expected.

\medskip

Let us note that one get the measure \ref{su2mes} directly by a
gauge fixing procedure without having to appeal to the general
theorem \ref{defmeasure}. Let
$g_i=\mat{a_i}{b_i}{-\bar{b}_i}{\bar{a}_i}$, $i=1,2$. The gauge
conditions we want to impose are $b_1(g_1)=0$ and $Im(b_2)(g_2)
=0$, $Im$ denotes the imaginary part. 
The Faddeev-Popov determinant is the determinant of the $3\times 3$ matrix 
$(Re(b_1)([X_i,g_1]); Im(b_1)([X_i,g_1]); Im(b_2)([X_i,g_2]))$,
 where $X_i,\,i=1,\cdots,3$ is a basis of $su(2)$.
This determinant is proportional to $(a_1-(a_1)^{-1})^2 b_{2}€$, it should 
multiply
the gauge fixed measure $dg_1 dg_2 \delta^2(b_1)\delta(Im(b_2))$, a direct
computation leads to the results \ref{su2mes}.

\subsection{$SL(2,\R)$}

In the case of $\slr$, the constraint reads
$\vec{X}(g_{1},g_{2}) \in J_{3}$ where
\beq
\label{j3}
J_{3} \equiv \{ X_{i}\in \R, \,
\Theta(\vec{X}) \neq 0,
(X_{1}^{2}\neq 1 \textrm{ or } X_{2}^2 \neq 1)  \} \textrm{ and } 
\vec{X} \neq I_{3} \}.
\eeq
The invariant measure is given by
\beq\label{slmes}
\int_{J_{3}}d\vec{X}.
\eeq
This result can be obtain using both
previous methods, gauge fixing or application of the
general formulas in the context of $\slr$.
It is interesting to note that $J_{3}$ correspond to a real section 
of $\C^{3}$ which is complementary to $I_{3}$ describing $\su$.
As in the case of $\su$, one can give a geometrical interpretation
of the configuration space $J_3$ and the non-degeneracy condition.
This will lead to a nice understanding of the singularity properties.

\medskip

It is well known that $\slr$ is isomorphic to $ AdS_3$,
the Anti-de-Sitter space in three dimensions, which can
be described as an hyperboloid in flat four
dimension space,
$AdS_3=\{-(X_0)^2+(X_1)^2+(X_2)^2-(X_3)^2 =-1\}$,
the isomorphism being
\beq \label{gx}
 g(X)=
\mat{X_0 +X_1} {X_2 +X_3}{X_2-X_3}{X_0-X_1}.
\eeq
$AdS_3$ is a
Lorentzian space and $SO(2,2)$ is its isometry group.
Then the space of couples of group elements $(g_1,g_2)$
corresponds to the space of geodesic triangles in $AdS_3$ with one
vertex fixed to be the identity. The adjoint action of $\slr$ on
$(g_1,g_2)$ translates into the action of the subgroup of
$SO(2,2)$ which fixes the identity,
hence into the action of the Lorentz group $SO(2,1)$
which rotates the triangles. So the space of orbits is the space
which describes
the intrinsic geometry of Anti-de-Sitter triangles. Such
triangles can be of four types~: they can be space-like, time-like,
null or degenerate (meaning that the three vertices of the triangle
belong to the same geodesic),
depending on whether they lay in a space-like,
time-like or null plane. The edges
of the triangles can also be of four types~: they can be time-like,
space-like, null or degenerate (meaning that the two vertices of the
edge coincide). Unlike the $SU(2)$ case,
the invariant geometry cannot be
fully characterized by the edge lengths since the length (more
precisely the square length) is zero for both a null edge and a
degenerate edge.
However, the following proposition \ref{tetriang} shows
that if we restrict the space of triangles
to triangles which satisfy the
condition $\Theta\neq 0$ then the geometry of the triangle is
uniquely determined by the lengths of its edges. The geometrical
meaning of this condition is the following:

\begin{prop}
\label{tetriang}
The condition $\Theta(g_1,g_2)= 0$ is equivalent to the condition
that the AdS triangle $(1,g_1,g_2)$ is either null or degenerate.
Moreover $\Theta(g_1,g_2) < 0$ (resp. $\Theta(g_1,g_2) > 0$) iff the
AdS triangle $(1,g_1,g_2)$ is spacelike (resp. timelike).
\end{prop}

In order to prove this proposition, we need to do some $AdS$ geometry.
It is convenient to consider $AdS$ space as embedded in the
projective space $\R^+P^3$, which is the space of half lines in
$\R^4$~: $AdS_3= \{(X_0:X_1:X_2:X_3)\in
\R^+P^3|\, -(X_0)^2 +(X_1)^2+(X_2)^2-(X_3)^2<0\} $.
The advantages
of such a representation of the $AdS$ space is
to simplify the geodesic geometry of $AdS$.
First the geodesics of
$AdS$ are the straight lines of $\R^+P^3$.
Moreover the geodesic
planes of $AdS_3$ are the intersection of $\R^+P^3$ planes with
$AdS$, they are therefore given by linear equations
$P_{(Y_0:Y_1:Y_2:Y_3)}=\{(X_0:X_1:X_2:X_3)\in \R^+P^3|\, X_iY^j
=0\}$ (the indices are raised using a lorentzian $(-,+,+,-)$
metric). Thus the geodesic hyperplanes of AdS are in one to one
correspondence with points of $R^+P^3$. Geometrically this means
that all the geodesics orthogonal to a given plane meet in one
point.
If the plane $P_{Y}$ is spacelike then $Y\cdot Y <0$ and
the refocusing point is in  AdS. This corresponds to the
attractive nature of negative cosmological constant where all
time-like geodesics re-focuse in a finite proper time. If the
plane $P_{Y}$ is time-like then $Y\cdot Y >0$. And if the plane
$P_{Y}$ is null then $Y\cdot Y =0$.
Moreover, in this latter case, $P_Y $ is tangent
to the quadric $Y\cdot Y =0$. \\
Next, we identify $\R^+P^3$
with the space of $2\times 2$ matrices modulo multiplication by a
positive scalar using $Y\in \R^+P^3 \rightarrow g(Y)$ as
in \ref{gx}. Now let's
consider the triangle $(1,g_1,g_2)$ and suppose that it is
non-degenerate i-e $[g_1,g_2]=g_1g_2-g_2 g_1\neq 0$.
Let's denote by
$Y(g_1,g_2)$ the $\R^+P^3$ element satisfying $g(Y(g_1,g_2))=
[g_1,g_2]$. It is clear that $P_{Y(g_1,g_2)}$ is the plane of the
triangle $(1,g_1,g_2)$ since $tr([g_1,g_2]g)=0$ if
$g= 1,g_1$ or $g_2$. Then a straightforward
computation shows that
\beq
 Y(g_1,g_2)\cdot Y(g_1,g_2)\equiv det([g_1,g_2])
 = 8\Theta(g_1,g_2),
 \eeq
and leads to the conclusion of the proposition \ref{tetriang}.

This proposition tells us that the space $J_3$ is the space of
non-degenerate and non-null triangles.
This space is disconnected,
and has two disconnected regions depending on
whether the normal
to the triangle is timelike or spacelike. There is no natural
distinction between past and future in $AdS_3$ since it
is periodic in
time and a timelike geodesic will come back to the initial normal
surface. Therefore we can define two Hilbert space structures, one
for the functions with support on spacelike triangles and one for the
functions with support on timelike triangles, with the scalar product
defined by the measure \ref{slmes}. The situation is however
drastically different from the one-loop case where the invariant
space was also disconnected but in that case the centralizer group
was drastically different in both regions.
In the present two petals case, we
see that the centralizer
i-e the group which fixes a given triangle, is
trivial in both sectors. This means that there is no superselection
rule avoiding to construct invariant operators mapping one sector
to another. Indeed one realizes that, even when we extend
the space $J_3$ to the space $\tilde{J}_3 $ of all
non-degenerate triangles (allowing null cases),
the centralizer of any triangle of $\tilde{J}_3 $  is still trivial.
Therefore we can extend the definition of the measure to
$\tilde{J}_3 $ which is {\bf connected} and
there exists one {\bf unique invariant measure} on this space.
In other words, we see that if $\phi(g_1,g_2)$ is a
function with compact support on $\tilde{J}_3 $ then
$^G\phi(g_1,g_2)=\int \phi(gg_1g^{-1},gg_2g^{-1})dg$ is well
defined for all $(g_1,g_2)\in \tilde{J}_3 $. This means that the
invariant distributions on $J_3$ obtained by group averaging can be
extended to invariant distributions on $\tilde{J}_3$. We expect
the spin network functionals to be of this type and therefore the
Hilbert space structure to be uniquely fixed in that case.

\section{The Hilbert Space of Spin Networks}
\label{Hilbert}
\subs{The Compact Group Case: The Ashtekar-Lewandowski construction}

 The Ashtekar-Lewandowski approach consists in the use
projective techniques in the {\it compact} group case to define
the space of generalized connections, a space of continuous
functions upon it (cylindrical functions) and a measure called the
Ashtekar-Lewandowski (AL) measure \cite{al4} which endow this
space with a Natural and difeomorphism invariant Hilbert space
structure. For a recent and complete review of this approach, one
can look at \cite{thiemann}.

First, we define a space of gauge invariant ``connections''
for each graph $\Gamma$, embedded in a spacelike manifold.
\beq
A_{\Gamma}=G^{\otimes E}/G^{\otimes V} =
\left\{ \left[
(g_{e_1},\dots,g_{e_E})\right] _{G^{\otimes V}} \right\} =\left\{
\{ (k^{-1}_{s(e_i)}g_{e_i}k_{t(e_i)},i=1\dots E),k_v \in G \}
\right\}
\eeq

We define a partial order $\prec$ over the set of graphs:
$\Gamma_1 \prec \Gamma_2$ iff $\Gamma_1$ can be obtained from
$\Gamma_2$ by removing edges and bivalent vertices. We then define
projections $p_{\Gamma_2 \Gamma_1}:A_{\Gamma_2}\arr A_{\Gamma_1}$
for $\Gamma_1\prec \Gamma_2$, by removing the extra edges and
contracting the extra bivalents vertices:

\beq
\left\{
\begin{array}{cccc}
\tr{removing the edge $i$} &
(g_1,\dots,g_i,\dots,g_E) & \arr
& (g_1,\dots,g_{i-1},g_{i+1},\dots,g_E) \\
\tr{bivalent vertex between $1$ and $2$} &
(g_1,g_2,\dots,g_E)
& \arr
& (g_1g_2^\epsilon,\dots,g_E)
\end{array}
\right.
\eeq
with $\epsilon=\pm 1$ depending on the relative
orientation on $g_1$ and $g_2$.

\medskip

Let's illustrate these rules with the example
of the reduction of the $\Theta$ graph
to a single loop:

\beq
\begin{array}{ccc}
\CBOX[1]{\FIGThetaNet{1}{2}{3}} & \longrightarrow &
\CBOX[1]{\looptwo{1}{3}} \\
(g_1,g_2,g_3)\sim (h^{-1}g_1k,h^{-1}g_2k,h^{-1}g_3k)
& \longrightarrow & (g_1,g_3) \sim (h^{-1}g_1k,h^{-1}g_3k) \\
\\
\CBOX[1]{\looptwo{1}{3}} & \longrightarrow &
\CBOX[1]{\loopone{1}} \\
(g_1,g_3) \sim (h^{-1}g_1k,h^{-1}g_3k) &
\longrightarrow &
G_1=g_1g_3^{-1}\sim h^{-1}G_1h
\end{array}
\eeq

Then, we can define the projective limit $\bar{A}$ as the set of
families of elements of $A_{\Gamma}$ consistent with the projections:

\beq
\bA=\left\{
(a_{\Gamma})_{\Gamma \mathrm{ graph}}
\in \times_{\Gamma} A_{\Gamma} \, / \,
\forall \, \Gamma_{1,2} \, , \,
\Gamma_1\prec \Gamma_2 \Rightarrow
p_{\Gamma_2 \Gamma_1}a_{\Gamma_2}=a_{\Gamma_1}
\right\}
\eeq

In the case of a compact group $G$,
the spaces $A_{\Gamma}$ are topological, compact and Haussdorf and the
projections are continuous, therefore $\bA$ with the Tychonov
topology (product topology) is compact and Haussdorf.
We can now construct continuous function on $\bA$.
We start by defining the spaces:

\beq
C^0(A_{\Gamma})=\left\{
f \in {\cal F}(A_{\Gamma},{\mathbf C}), f\tr{ continuous}
\right\}
\eeq
The projections $p$ induce some injections between the spaces of functions
$C^0(A_{\Gamma_1})$ and $C^0(A_{\Gamma_2})$ for $\Gamma_1 \prec \Gamma_2$:

\beqs
i_{\Gamma_1\Gamma_2}:  C^0(A_{\Gamma_1}) & \arr & C^0(A_{\Gamma_2})\\
\phi(\{g_e\}_{e\in\Gamma_1})
& \rightarrow & \tl\phi(\{g_e\}_{e\in\Gamma_2})=
\phi(p_{\Gamma_2\Gamma_1}\{g_e\}_{e\in\Gamma_2})
\eeqs
We define the following equivalence relation:

\beq
\begin{array}{ccc}
f_{\Gamma_1}\in C^0(A_{\Gamma_1}) \sim
f_{\Gamma_2}\in C^0(A_{\Gamma_2})
& \Leftrightarrow &
\exists \, \Gamma_3 \succ \Gamma_1 , \Gamma_2, \,
i_{\Gamma_1\Gamma_3}f_{\Gamma_1}=
i_{\Gamma_2\Gamma_3}f_{\Gamma_2} \\
& \Leftrightarrow &
\forall \, \Gamma_3 \succ \Gamma_1 , \Gamma_2, \,
i_{\Gamma_1\Gamma_3}f_{\Gamma_1}=
i_{\Gamma_2\Gamma_3}f_{\Gamma_2}
\end{array}
\eeq
This allows us to define the space of Cylindrical functions:

\beq
Cyl(\bA)=\bigcup_{\Gamma}C^0(A_{\Gamma}) {\Big /} \sim
\eeq
We divide by the previous equivalence relation in order to remove
the redundancies due to the existence of the injections.
On $Cyl(\bA)$, we can define a norm

\begin{equation}
\| [ f_{\Gamma} ]_\sim \| = \sup_{x_\Gamma \in A_\Gamma} |f_\Gamma (x_\Gamma)|
\end{equation}
Then the completed space
is a abelian $C^*$ algebra, to which we can apply the Gelfand-Naimark theorem.
It states that it is the algebra of continuous functions on
a certain {\bf compact Haussdorf} space called the Gelfand spectrum of the
 $C^*$ algebra. In \cite{al4}, Ashtekar and Lewandowski prove
that its Gelfand spectrum is simply $\bA$ i.e that we have the following
isomorphism:

\beq
Cyl(\bA)\approx C^0(\bA)
\eeq
Choosing measures $\dmu^{(\Gamma)}$
-the Haar measure- on the spaces of discrete connections $A_{\Gamma}$
and checking that
they are consistent with the injections

\beq
\forall \, \Gamma_1 \prec \Gamma_2 \, ,
\, i_{\Gamma_1\Gamma_2}\dmu^{(\Gamma_2)}=\dmu^{(\Gamma_1)}
\eeq
we can define a measure $\overline{\tr{d}\mu}$
- the {\it Ashtekar-Lewandowski} measure- on $\bA$
by considering their projective limit.
And our final Hilbert
space will be $\H_{\mathrm{cyl}}=L^2(\bar{A},\overline{\tr{d}\mu})$.

\subs{An Alternative: the GNS construction}

An elegant way of constructing the Hilbert space $\H_{\mathrm{cyl}}$
is using the GNS (Gelfand-Naimark-Segal) construction \cite{al5,gns}.
One considers the algebra ${\cal A}$ of all cylindrical functions
$f_\Gamma$ (on all graphs $\Gamma$) with the normal multiplication
law between functions. One defines the norm sup on this space as in
the previous paragraph:

\beq
\| f_\Gamma \| =\sup_{A_\Gamma}|f_\Gamma|
\eeq
One can then complete ${\cal A}$ to a $C^*$ algebra $\bar{{\cal A}}$.
On $\bar{{\cal A}}$, we define a {\it state} $\om$- a positive linear form -
which is simply the integration:

\beq
\om(f_\Gamma)=\int_{A_\Gamma} \dmu^{(\Gamma)} f_\Gamma
=\int_{SU(2)^E}\tr{d}g_1 \dots\tr{d}g_E f_\Gamma(g_1,\dots,g_E)
\eeq
$\om$ induced an inner product
$\langle f_{\Gamma_1}|f_{\Gamma_2} \rangle=\om(f^*_{\Gamma_1}f_{\Gamma_2})$
remembering that the product of the two cylindrical functions
is a cylindrical function based on any graph bigger than both $\Gamma_1$ and
$\Gamma_2$. We then define the Gelfand ideal

\beq
{\cal I}=\{ a\in \bar{{\cal A}} | \om(a^*a)=0\}
\eeq
We get a positive definite scalar product on the space
${\cal H}_{\mathrm{gns}}=\bar{{\cal A}}/{\cal I}$. And we get the physical
Hilbert space by completing this space to $\overline{{\cal H}_{\mathrm{gns}}}$.
It is straightforward to check that the equivalence relation $\sim$ is
the same as defined by ${\cal I}$ so that
$\overline{{\cal H}_{\mathrm{gns}}}=\H_{\mathrm{cyl}}$.

\medskip
Let's make this construction explicit using the spaces
$\H_\Gamma=L^2(A_\Gamma,\dmu^{(\Gamma)})=L^2(A_\Gamma,\dmu^E)=$
with $\dmu$ being the Haar measure on $SU(2)$. $\H_\Gamma$ is the
Hilbert space of spin networks based on the graph $\Gamma$. The
usual basis is indeed the spin networks basis. These spin networks
are labelled by irreducible representations $j$ of $SU(2)$ on each
edge and intertwiners $i$ for each vertex. Then, the function is
defined by taking the group elements in the edge representations
and contracting them using the intertwiners.

Let's give a decomposition of $\H_\Gamma$ on which it will be
easy to implement the equivalence relation $\sim$. If an edge $e$
of $\Gamma$ is labelled by the representation $j=0$, then the
corresponding spin network function will not depend on the group
element $g_e$: it will be equal to the spin network defined on
$\Gamma'=\Gamma\setminus\{e\}$ with the same labels. So we can
decompose $\H_\Gamma$ into the direct sum of Hilbert spaces
$\tl{\H}_{\Gamma'}$, $\Gamma'\subset\Gamma$, of spin networks
based on $\Gamma'$ {\bf with no trivial representations $j=0$}.
Furthermore, if we consider an arbitrary graph $\Gamma_1$ (which
is not  the single loop with a single vertex) and a graph
$\Gamma_2$ obtained by removing a bivalent vertex from $\Gamma_1$,
the space $\tl{\H}_{\Gamma_1}$ and $\tl{\H}_{\Gamma_2}$ are
isomorphic using the restriction of the injection
$i_{\Gamma_2\Gamma_1}$ to $\tl{\H}_{\Gamma_2}$.
  This means that we can decompose $\H_\Gamma$ as the direct sum of spaces
$\H_\Gamma'$ with $\Gamma'\subset \Gamma $ containing no bivalent
vertex (including loops with no vertex at all). To sum up this, we
define $\G$ the set of all graphs and $\Gtl$ {\it the set of all
graphs which don't contain bivalent vertices}. We have:

\beq
\H_\Gamma=\bigoplus_{\Gamma' \in \Gtl,\,
\Gamma'\subset\Gamma}(i_{\Gamma'\Gamma}) \Htl_{\Gamma'}
\eeq
Using these new notations, we have
$\H_{\mathrm{gns}}=
\bigoplus_{\Gamma \in {\Gtl}}\Htl_{\Gamma} $
which implements in a practical way the non-direct sum
$+_{\Gamma \in {\Gtl}}\H_{\Gamma}\equiv
\bigoplus_{\Gamma \in {\G}}\H_{\Gamma}/\sim$.

\subs{The Non-Compact Group Case}

Let's now assume the group $G$ is non-compact.
The obstacle to applying the AL
construction is the non-compactness of the $A_\Gamma$ spaces.
There is no problem defining the projections
and injections.
However, the space $\bar{A}$ is non-compact and
therefore we can not obtain it as Gelfand spectrum.
Moreover, we can not define a norm $\| f_\Gamma \|$
on the spaces of continuous functions $C^0(A_\Gamma)$ so
that $Cyl(\bar{A})$ does not have any norm and
can not be completed into a $C^*$ algebra.
And finally, the family of measures
$\dmu^{(\Gamma)}$ is not consistent
with the partial order on the set of graphs.
To save some results of the AL approach, one could try to
compactify the spaces $A_\Gamma$ or to impose some cut-off on the
group. Nevertheless, this makes it hard to deal with the gauge
invariance. One could also change the definition of
$C^0(A_\Gamma)$ by taking the bounded continuous functions in
order to define a norm on these spaces. Nevertheless, it is not
clear what would be its Gelfand spectrum.

One promising approach would be to use the fact that $A_\Gamma$ is
an algebraic space. It can therefore be recovered  not as a
Gelfand spectrum but as an algebraic spectrum of the affine
algebra $P(A_\Gamma)$ of polynomial function. The problem is that
the union of all such affine algebras modulo $\sim$ in no longer
finitely generated so the usual theorems of algebraic geometry
can not applied and it is not clear if one can define an algebraic
dual of that space. But we still think that this road is worth
pursuing.

Here, we choose to concentrate on defining a Hilbert space
-the Hilbert space of spin networks- and we don't tackle the problem of
constructing it as a $L^2$ space. Our construction will be based on the results
obtained from the GNS approach; in particular, we won't need
the projections/injections structure.
The drawback of this approach is that we don't construct  the
space of generalized connections $\bA$. So we can not interpret our
Hilbert space as a $L^2$ space: we lose
some aspects of the ``wave function''
interpretation. But for all practical purpose the Hilbert space
structure is all of what we need.

\medskip

So, what is the structure we are left with? In the non-compact
group case the trivial representation $j=0$ is not a $L^2$
representation. Any function not depending on a group element is
clearly not normalizable. In other words the trivial
representation doesn't appear in the decomposition of
$L^2(A_\Gamma)$. This mean that we have built directly the spaces
$\Htl$ defined in the previous paragraph. And we build the
configuration space as a direct sum of these spaces:

\beq 
\label{Hfinal}
\H_{\mathrm{config}}= \bigoplus_{\Gamma
\in\Gtl}\Htl_{\Gamma}
\eeq

There is a possible normalization ambiguity in the above
summation. A priori, we are free to normalize the different
$\Htl_{\Gamma}$ spaces as we wish. This relative normalization of
the Hilbert spaces can be traced down to an ambiguity in the
definition of the Haar measure used to define the measure
$\dmu^{(\Gamma)}$ of each Hilbert space. In the compact case, we
fix these measures to be probability measures \cite{baez} and we
normalize the Haar measure such that the group gets a unit volume.
This makes the measures consistent with the projection structure of
the Ashtekar-Lewandowski construction. In the non-compact group
case, it is impossible to define such a normalization. However,
looking at the way the Haar measure comes into the definition of
\Ref{defdmun} the measures over different graphs, it is natural to
require that the Haar measure be normalized the same way for all
measures. More precisely, if we take an integrable function over
$G^{(n+1)}/Ad(G)$, we can integrate out one of its variable using
the Haar measure, and we would get an integrable function over
$G^n/Ad(G)$. Then, it is natural to require that the integrals of
the two functions be equal.

This argument fixes the Haar measure up to a constant. And if we rescale the
Haar measure by a factor $\alpha$, then the measures $\dmu_n$ are to be scaled
by $\alpha^{(n-1)}$. And we can think of the normalization of the Haar measure
as the choice of a scale in our physical theory.

\medskip

Now the space $\H_{\mathrm{config}}$ defined in \Ref{Hfinal}
doesn't seem to be a $L^2$ space. Nevertheless, it carries some
Fock space structure. In that frame, the projection/injection structure of
the AL approach would be replaced by creation and annihilation
operators. These would act like isometries between the different
Hilbert spaces $\Htl_{\Gamma}= L^2(G^E/G^V)=L^2(G^{h_\Gamma}/Ad(G))$
and could fix the
normalization ambiguity. More precisely, let's consider an infinite
graph $\Gamma_\infty$ i-e a sequence of graphs $(\Gamma_i)_{i\in
N},\, \Gamma_i \in  \Gtl$ such that $\Gamma_i \prec \Gamma_{i+1}$
and the inclusion is strict. Then the space
\beq
{\cal{F}}_{\Gamma_\infty} =\bigoplus_{i}\Htl_{\Gamma_i},
\eeq
looks like a Fock space where the addition of a loop would be a creation
operator.

\medskip

The difficulty of endowing ${\cal{F}}_{\Gamma_\infty}$ of
a Fock space structure comes from the residual $Ad(G)$
non-compact gauge symmetry. There exists a natural
gauge fixing through the possibility
of erasing this remaining symmetry by considering cylindrical
functions which are gauge invariant
but at a single vertex of the graph.
Indeed, following the gauge fixing procedure
described in section \ref{treefixing} based at
the point $A$, the space of such graph connections is
simply $G^{\otimes h_\Gamma}$ and the corresponding
Hilbert space of states is
$L^2(G^{\otimes h_\Gamma},dg^{\otimes h_\Gamma})$.
It is then possible to pile these spaces into
a Fock space of states ${\cal F}$ by carefully
summing over graphs. The connection states can
be seen as a set of loops whose base point is $A$
(flower around the vertex $A$).
The creation and annihilation operators
acting as usually to go from
$L^2(G^N)$ to $L^2(G^{N\pm 1})$ then create
or destroy a loop from $A$.
Thus, from this point of view,
${\cal F}$ represents the fluctuations
of the connection around the point $A$. Then, what
about the gauge invariance at the point $A$?
Imposing it directly on ${\cal F}$ leads to divergence problems.
Nevertheless, instead of imposing gauge invariance, we could
place ourself at the point $A$
and ignore the gauge invariance but
instead impose that the considered states transform nicely
under $G$ and belong to a given
representation of the group $G$.
However this means introducing
by hand in the theory an
observer at the point $A$, represented by the chosen representation.
And in the present work, we
prefer to tackle the issue of considering fully gauge invariant
functionals and study the sum of the spaces
$L^2(G^{\otimes h_\Gamma}/Ad(G))$.

\subsection{Towards a Fock space for the space of connections}

\label{section:particle}

%\sss{The particle analogy}

We are interested by gluing together $L^2(G^n/Ad(G))$ spaces. An
useful analogy is interpreting these spaces as state spaces of
particles living on the group $G$ \cite{particle}. Indeed,
$L^2(G^n,({\textrm{d}}g)^n)$ - $\textrm{d}g$ is the Haar measure
on $G$ - is the space corresponding to a free particle living on
the group $G^n$ or equivalently $n$ free particles living on the
group $G$. Its action evaluated on a function
$g(t):\Rnum\rightarrow G^n$ is

\beq
S_{\mathrm{free}}=\f{1}{2}\int dt \textrm{Tr}
\left((g^{-1}\partial_tg)^2\right)
\eeq
One can check this action is invariant under (constant)
left and right multiplication in $G^n$.
We can do the hamiltonian analysis of this system and the phase space is
the tangent bundle of the group $G$.
The equation of motion is
\beq
\pp_t(g^{-1}\pp_tg)=0
\eeq
We choose $\pi^{(l)}=g^{-1}\pp_tg$ as momentum
(instead of the canonical momentum),
it is the Noether charge associated to the left invariance.
The solutions are then parameterized as

\beq
g(t,g_0)=g_0 \exp(\pi^{(l)} t)
\eeq
We could also choose the right momentum
defined by $\pi^{(r)}=-\pp_tg g^{-1}$ and then
the solutions would be
\beq
g(t,g_0)=\exp(-\pi^{(r)}t)g_0
\eeq
which are the geodesics.

The Poisson bracket reads
\beq
\begin{array}{ccc}
\{g,\hat{g}\} &= & 0 \\
\{g,\pi^{(l)}_X\} &=& Xg\\
\{\pi^{(l)}_X,\pi^{(l)}_Y\} &=& \pi^{(l)}_{[X,Y]}
\end{array}
\eeq
where $g,\, \hat{g}$ are group elements, $X,Y$ are in the Lie algebra,
and $\pi^{(l)}_X=\textrm{Tr}(X \pi^{(l)})$ is the
component of $\pi^{(l)}$ in the $X$ direction.

\medskip
One can create a Fock space for the free particles states

\beq
{\cal F}=\bigoplus_{n\ge 0} L^2(G^n)
\eeq
We can construct creation operators $a^\dagger_\varphi$
and annihilation operators $a_\varphi$ which are adding or
removing a one particle state to a (symmetrized) $n$ particle state
-let's call it $\psi$:

\beq
\label{dest}
(a_\varphi\psi)(g_1,g_2,\dots,g_{n-1})=
\int dg_n \psi(g_1,g_2,\dots,g_n)\overline{\varphi}(g_n)
\eeq

\beq \label{crea}
(a^\dagger_\varphi\psi)(g_1,g_2,\dots,g_{n+1})=\sum_i
\psi(g_1,\dots, g_{i-1},g_{i+1},\dots,g_n)\varphi(g_{i})
\eeq

We can also write a (free) field theory corresponding
to this Fock space.
Indeed, Let us define a field operator $\Phi(g) = a_{\delta_{g}} $, 
$\Phi^\dagger(g) = a^\dagger_{\delta_{g}} $, where $\delta_{g}$ denote 
the Dirac delta function supported at $g$.
Then, the action of the total impulsion operator on the Hilbert space of N 
particles can be written in terms of the field operators
$ \sum_{i=1}^{N}\pi^{i{(l/r)}}_{X} = 
\int_{G}Ä dg \Phi^\dagger(g)(-i\nabla^{{(l/r)}}_{X})\Phi(g)$, where $ \nabla^{{(l/r)}}_{X}$
denote the left or right invariant derivative operator in the 
direction of $X$. In the same way the hamiltonian operator can be 
written as 
\beq
H= -\int_{G} dg \Phi^\dagger(g)\Delta \Phi(g), 
\eeq 
and the action governing the quantization and the dynamic of the field is 
expressed in term 
of a space time field $\Phi(t,g)$:
\beq \label{Sfree}
S[\Phi(t,g)]=\int_{\R\times G}dt dg 
\Phi^\dagger(g)(i{\partial \over \partial t} + \Delta)\Phi(g).
\eeq

We now wish to follow the same steps for gauged particles
i.e in the case that we gauge the global $Ad(G)$ symmetry.
This can be achieved by introducing a gauge fields $A$
living in the Lie algebra $\G$, and the action reads (we have
slightly modified the action given in \cite{particle}):

\beqs
S_{\mathrm{gauged}} [g(t)\in G^n,A]&= & -\f{1}{2}\int dt
\textrm{Tr}\left((g^{-1}\partial_tg)^2\right) \nonumber \\
&& +\int dt \textrm{Tr}\left(
(g\pp_tg^{-1})A+A(g^{-1}\pp_tg)+gAg^{-1}A-A^2
\right)
\eeqs
This action is invariant under the following $Ad(G)$ gauge invariance for
arbitrary $G$-valued $h(t)$:

\beq
\left\{
\begin{array}{ccc}
g & \rightarrow & hgh^{-1}\\
A & \rightarrow & hA h^{-1} + h\pp_th^{-1}
\end{array}
\right.
\eeq
 The space of states of our system will be $L^2(G^n/Ad(G))$.
For these gauged particles,
we would like to do the same thing as for the free particles
i.e write down creation and annihilation operators and a
corresponding field theory.
The problem is the change of symmetry:
the symmetry in $L^2(G^n/Ad(G))$ is a global symmetry $Ad(G)$ on
the system of $n$ particles and it is hard to have a Fock space 
interpretation.
An analogy would be to study a system of $N$ particles in space-time
which would be invariant under global
Poincar{\'e} transformations.

The easiest way to write creation and annihilation
operators would be through gauge fixing.
Starting from a graph $\Gamma$ and going through the gauge fixing procedure,
we have seen that the space $L^2(A_\Gamma)$ is naturally isomorphic to
$L^2(G^n)$.
Therefore the space associated with an infinite graph is similar Fock space.
And in this context creation and annihilation operators are adding or removing
a loop to the graph.
We feel that it will be interesting to have a deeper understanding of this 
ideas and of the field theory behind this. Note that the action of the field 
theory behind the gauged particle is obtain by introducing a gauge 
field $A(t)$ and a term to the action \Ref{Sfree}:
\beq
\int dt dg A(t)( \nabla^{{(l)}} - \nabla^{{(r)}})\Phi(g,t).
\eeq

\section*{Conclusion}
In this paper, we have defined the notion of Spin network states
for non-compact reductive groups. We have shown how to construct
the quotient space of  graph connections as the algebraic dual of
a polynomial algebra. We have also constructed, by a careful gauge
fixing procedure, a canonical measure on this space which turned out
to be independent of
any gauge fixing choices. This measure defines a Hilbert space
structure for each graph, and spin networks states are defined as
generalized eigenvectors of invariant, hermitic differential
operators. We have explicitly realized  all these ideas in the
context of $\slr$ and $\slc$ by a direct analysis
of the quotient space
and measure in the simplest cases. Finally we have discussed the
nature of the full Hilbert space based on all graphs and we have
shown that a natural Fock structure appears in this context.

The work we have done is the first step toward a full comprehension 
of non-compact spin networks, i-e identical to the one we 
have for the compact
ones.  As we have stressed in our paper, an understanding of the
full Hilbert space as an $L^2$ space is still missing. We expect
that this should come together with an interpretation of the full
space of gauged connections as an algebraic dual.
Also, a more detailed and explicit study of the space of spin networks
would be interesting to pursue in order to reach a deeper
understanding of
their analytic  properties, in the spirit of the work of
Harish-Chandra on characters of non-compact groups. 
Finally, we
feel that the Fock space structure which is emerging in our
construction is something important that should be put on firmer
basis. Nevertheless, this work opens the possibility to study
the Hilbert spaces of non-compact spin networks that arises in
Lorentzian formulations of gravity and allows us to discuss the
spectra of geometrical operators in this context \cite{crl}.

{\bf Acknowledgment} L.F. would like to thank F. Delduc for a 
discussion.
This work has been supported by an ACI-blanche grant.

\end{document}

%% file: flowerT.eepic
\setlength{\unitlength}{0.00066667in}
\begingroup\makeatletter\ifx\SetFigFont\undefined%
\gdef\SetFigFont#1#2#3#4#5{%
  \reset@font\fontsize{#1}{#2pt}%
  \fontfamily{#3}\fontseries{#4}\fontshape{#5}%
  \selectfont}%
\fi\endgroup%
{\renewcommand{\dashlinestretch}{30}
\begin{picture}(2579,1537)(0,-10)
\path(957,750)(955,750)(951,751)
	(943,752)(931,754)(915,757)
	(896,761)(874,765)(850,771)
	(824,777)(797,784)(770,792)
	(742,802)(715,812)(687,825)
	(660,839)(632,856)(605,876)
	(578,898)(554,923)(533,949)
	(515,976)(501,1000)(490,1023)
	(482,1042)(477,1059)(473,1074)
	(470,1087)(468,1099)(467,1110)
	(467,1121)(467,1133)(467,1145)
	(468,1159)(469,1174)(471,1192)
	(474,1210)(479,1230)(486,1250)
	(496,1269)(507,1282)(519,1293)
	(531,1302)(544,1310)(556,1316)
	(567,1322)(577,1326)(587,1330)
	(596,1333)(605,1336)(613,1339)
	(621,1341)(629,1343)(637,1344)
	(646,1345)(656,1344)(668,1343)
	(680,1340)(694,1335)(710,1328)
	(727,1318)(745,1305)(765,1289)
	(784,1269)(804,1244)(823,1215)
	(840,1185)(855,1154)(869,1123)
	(880,1091)(891,1059)(901,1026)
	(909,993)(917,961)(925,928)
	(931,897)(937,867)(942,840)
	(946,815)(950,794)(953,777)
	(955,765)(956,757)(957,752)(957,750)
\path(957,750)(957,752)(958,756)
	(959,764)(961,776)(964,792)
	(968,811)(972,833)(978,858)
	(984,883)(991,910)(999,938)
	(1008,965)(1019,993)(1031,1020)
	(1046,1048)(1062,1076)(1082,1103)
	(1104,1129)(1129,1154)(1155,1175)
	(1182,1193)(1206,1207)(1228,1218)
	(1248,1226)(1265,1231)(1280,1235)
	(1293,1238)(1305,1240)(1316,1241)
	(1327,1241)(1339,1241)(1351,1241)
	(1365,1240)(1380,1239)(1398,1237)
	(1416,1234)(1436,1229)(1456,1222)
	(1475,1212)(1488,1201)(1499,1189)
	(1508,1177)(1516,1164)(1522,1152)
	(1528,1141)(1532,1130)(1536,1120)
	(1539,1111)(1542,1103)(1545,1095)
	(1547,1087)(1549,1079)(1550,1070)
	(1550,1061)(1550,1051)(1548,1040)
	(1545,1027)(1541,1013)(1533,997)
	(1524,980)(1511,962)(1495,942)
	(1475,923)(1450,903)(1421,884)
	(1392,867)(1361,852)(1329,838)
	(1297,826)(1265,816)(1232,806)
	(1200,797)(1167,790)(1135,782)
	(1104,776)(1074,770)(1046,765)
	(1022,761)(1001,757)(984,754)
	(972,752)(964,751)(959,750)(957,750)
\path(957,750)(957,748)(956,744)
	(955,736)(953,724)(950,708)
	(946,689)(942,667)(936,642)
	(930,617)(923,590)(915,562)
	(905,535)(895,507)(882,480)
	(868,452)(851,424)(831,397)
	(809,371)(784,346)(758,325)
	(731,307)(707,293)(684,282)
	(665,274)(648,269)(633,265)
	(620,262)(608,260)(597,259)
	(586,259)(574,259)(562,259)
	(548,260)(533,261)(515,263)
	(497,266)(477,271)(457,278)
	(438,288)(425,299)(414,311)
	(405,323)(397,336)(391,348)
	(385,359)(381,370)(377,380)
	(374,389)(371,397)(368,405)
	(366,413)(364,421)(363,430)
	(362,439)(363,449)(364,460)
	(367,473)(372,487)(379,503)
	(389,520)(402,538)(418,558)
	(438,577)(463,597)(492,616)
	(522,633)(553,648)(584,662)
	(616,674)(648,684)(681,694)
	(714,703)(746,710)(779,718)
	(810,724)(840,730)(867,735)
	(892,739)(913,743)(930,746)
	(942,748)(950,749)(955,750)(957,750)
\path(957,750)(959,750)(963,749)
	(971,748)(983,746)(999,743)
	(1018,739)(1040,735)(1064,729)
	(1090,723)(1117,716)(1144,708)
	(1172,698)(1199,688)(1227,675)
	(1254,661)(1282,644)(1309,624)
	(1336,602)(1360,577)(1381,551)
	(1399,524)(1413,500)(1424,477)
	(1432,458)(1437,441)(1441,426)
	(1444,413)(1446,401)(1447,390)
	(1447,379)(1447,367)(1447,355)
	(1446,341)(1445,326)(1443,308)
	(1440,290)(1435,270)(1428,250)
	(1418,231)(1407,218)(1395,207)
	(1383,198)(1370,190)(1358,184)
	(1347,178)(1336,174)(1326,170)
	(1317,167)(1309,164)(1301,161)
	(1293,159)(1285,157)(1276,156)
	(1267,155)(1257,156)(1246,157)
	(1233,160)(1219,165)(1203,172)
	(1186,182)(1168,195)(1148,211)
	(1129,231)(1109,256)(1090,285)
	(1073,315)(1058,346)(1045,377)
	(1033,409)(1022,441)(1013,474)
	(1004,507)(996,539)(989,572)
	(983,603)(977,633)(972,660)
	(968,685)(964,706)(961,723)
	(959,735)(958,743)(957,748)(957,750)
\put(784,808){\makebox(0,0)[lb]{\smash{{{\SetFigFont{10}{12.0}{\familydefault}{\mddefault}{\updefault}C}}}}}
\put(1418,1327){\makebox(0,0)[lb]{\smash{{{\SetFigFont{10}{12.0}{\familydefault}{\mddefault}{\updefault}$G^{(T)}_9$}}}}}
\put(1533,288){\makebox(0,0)[lb]{\smash{{{\SetFigFont{10}{12.0}{\familydefault}{\mddefault}{\updefault}$G^{(T)}_7$}}}}}
\put(0,1125){\makebox(0,0)[lb]{\smash{{{\SetFigFont{10}{12.0}{\familydefault}{\mddefault}{\updefault}$G^{(T)}_1$}}}}}
\put(225,0){\makebox(0,0)[lb]{\smash{{{\SetFigFont{10}{12.0}{\familydefault}{\mddefault}{\updefault}$G^{(T)}_3$}}}}}
\end{picture}
}

%% file: move.eepic
\setlength{\unitlength}{0.00083333in}
\begingroup\makeatletter\ifx\SetFigFont\undefined%
\gdef\SetFigFont#1#2#3#4#5{%
  \reset@font\fontsize{#1}{#2pt}%
  \fontfamily{#3}\fontseries{#4}\fontshape{#5}%
  \selectfont}%
\fi\endgroup%
{\renewcommand{\dashlinestretch}{30}
\begin{picture}(3719,1275)(0,-10)
\thicklines
\put(675.000,457.500){\arc{615.000}{6.0619}{9.6461}}
\put(2925.000,457.500){\arc{615.000}{6.0619}{9.6461}}
\thinlines
\path(675,1050)(975,525)
\path(2925,1050)(2625,525)
\thicklines
\path(1575,600)(2250,600)
\blacken\thinlines
\path(2070.000,532.500)(2250.000,600.000)(2070.000,667.500)(2124.000,600.000)(2070.000,532.500)
\thicklines
\path(675,1050)(375,525)
\path(2925,1050)(3225,525)
\put(600,1125){\makebox(0,0)[lb]{\smash{{{\SetFigFont{10}{12.0}{\rmdefault}{\mddefault}{\updefault}$v$}}}}}
\put(2850,1125){\makebox(0,0)[lb]{\smash{{{\SetFigFont{10}{12.0}{\rmdefault}{\mddefault}{\updefault}$v$}}}}}
\put(975,675){\makebox(0,0)[lb]{\smash{{{\SetFigFont{10}{12.0}{\rmdefault}{\mddefault}{\updefault}$f\notin T$}}}}}
\put(3150,750){\makebox(0,0)[lb]{\smash{{{\SetFigFont{10}{12.0}{\rmdefault}{\mddefault}{\updefault}$f\in  U$}}}}}
\put(600,0){\makebox(0,0)[lb]{\smash{{{\SetFigFont{10}{12.0}{\rmdefault}{\mddefault}{\updefault}$\in T$}}}}}
\put(2850,0){\makebox(0,0)[lb]{\smash{{{\SetFigFont{10}{12.0}{\rmdefault}{\mddefault}{\updefault}$\in U$}}}}}
\put(2325,750){\makebox(0,0)[lb]{\smash{{{\SetFigFont{10}{12.0}{\rmdefault}{\mddefault}{\updefault}$e\notin U$}}}}}
\put(0,675){\makebox(0,0)[lb]{\smash{{{\SetFigFont{10}{12.0}{\rmdefault}{\mddefault}{\updefault}$e\in T$}}}}}
\end{picture}
}

%% file: move2.eepic
\setlength{\unitlength}{0.00058333in}
\begingroup\makeatletter\ifx\SetFigFont\undefined%
\gdef\SetFigFont#1#2#3#4#5{%
  \reset@font\fontsize{#1}{#2pt}%
  \fontfamily{#3}\fontseries{#4}\fontshape{#5}%
  \selectfont}%
\fi\endgroup%
{\renewcommand{\dashlinestretch}{30}
\begin{picture}(4948,3059)(0,-10)
\put(1362.500,2451.500){\arc{617.454}{5.9102}{8.4083}}
\path(1706.842,2435.937)(1650.000,2564.000)(1632.648,2424.967)
\path(1339.852,2180.479)(1200.000,2189.000)(1315.421,2109.570)
\put(2297.635,2609.608){\arc{697.751}{0.3530}{3.2259}}
\path(2585.979,2354.432)(2625.000,2489.000)(2522.165,2393.836)
\path(2002.081,2508.928)(1950.000,2639.000)(1927.534,2500.701)
\put(3053.572,2274.714){\arc{622.906}{1.5019}{3.6052}}
\path(2935.865,1947.485)(3075.000,1964.000)(2947.281,2021.611)
\path(2778.838,2273.941)(2775.000,2414.000)(2706.052,2292.027)
\path(1125,2639)(2175,2714)
\path(2175,2714)(3150,2414)
\path(3150,1589)(2325,1064)
\path(2325,1064)(1125,1289)
\thicklines
\path(1106,1298)(1629,44)
\path(1106,1298)(479,44)
\path(1106,2655)(1106,1298)
\dashline{120.000}(3150,2414)(3150,1589)
\put(0,1889){\makebox(0,0)[lb]{\smash{{{\SetFigFont{10}{12.0}{\rmdefault}{\mddefault}{\updefault}$e\in T\setminus U$}}}}}
\put(2625,2714){\makebox(0,0)[lb]{\smash{{{\SetFigFont{10}{12.0}{\rmdefault}{\mddefault}{\updefault}$f_2$}}}}}
\put(1275,2864){\makebox(0,0)[lb]{\smash{{{\SetFigFont{10}{12.0}{\rmdefault}{\mddefault}{\updefault}$f_1\in T\cap U$}}}}}
\put(3225,1964){\makebox(0,0)[lb]{\smash{{{\SetFigFont{10}{12.0}{\rmdefault}{\mddefault}{\updefault}$f\in U\setminus T$}}}}}
\end{picture}
}

%% file: theta.eepic
\setlength{\unitlength}{0.00041667in}
\begingroup\makeatletter\ifx\SetFigFont\undefined%
\gdef\SetFigFont#1#2#3#4#5{%
  \reset@font\fontsize{#1}{#2pt}%
  \fontfamily{#3}\fontseries{#4}\fontshape{#5}%
  \selectfont}%
\fi\endgroup%
{\renewcommand{\dashlinestretch}{30}
\begin{picture}(2359,2187)(0,-10)
\put(1315.795,1084.682){\arc{1406.508}{4.5441}{6.2714}}
\put(1315.795,1102.318){\arc{1406.507}{0.0118}{1.7391}}
\put(1080.939,1081.743){\arc{1412.058}{3.1575}{4.8790}}
\thicklines
\path(1015.926,1724.624)(1198.000,1778.000)(1020.315,1844.543)
\thinlines
\put(1080.939,1105.257){\arc{1412.057}{1.4042}{3.1256}}
\thicklines
\path(1020.315,342.457)(1198.000,409.000)(1015.926,462.376)
\thinlines
\put(356,1094){\blacken\ellipse{84}{84}}
\put(356,1094){\ellipse{84}{84}}
\put(2006,1094){\blacken\ellipse{84}{84}}
\put(2006,1094){\ellipse{84}{84}}
\path(375,1056)(1200,1056)
\thicklines
\path(1380.000,1116.000)(1200.000,1056.000)(1380.000,996.000)
\path(1200,1056)(2025,1056)
\put(0,981){\makebox(0,0)[lb]{\smash{{{\SetFigFont{9}{10.8}{\rmdefault}{\mddefault}{\updefault}A}}}}}
\put(2175,981){\makebox(0,0)[lb]{\smash{{{\SetFigFont{9}{10.8}{\rmdefault}{\mddefault}{\updefault}B}}}}}
\put(975,1956){\makebox(0,0)[lb]{\smash{{{\SetFigFont{9}{10.8}{\rmdefault}{\mddefault}{\updefault}$g_1$}}}}}
\put(1050,81){\makebox(0,0)[lb]{\smash{{{\SetFigFont{9}{10.8}{\rmdefault}{\mddefault}{\updefault}$g_2$}}}}}
\put(975,1281){\makebox(0,0)[lb]{\smash{{{\SetFigFont{9}{10.8}{\rmdefault}{\mddefault}{\updefault}$g_3$}}}}}
\end{picture}
}

%% file: 2petal.eepic
\setlength{\unitlength}{0.00041667in}
\begingroup\makeatletter\ifx\SetFigFont\undefined%
\gdef\SetFigFont#1#2#3#4#5{%
  \reset@font\fontsize{#1}{#2pt}%
  \fontfamily{#3}\fontseries{#4}\fontshape{#5}%
  \selectfont}%
\fi\endgroup%
{\renewcommand{\dashlinestretch}{30}
\begin{picture}(3319,1119)(0,-10)
\put(937,563){\ellipse{1068}{1068}}
\put(1987,563){\ellipse{1068}{1068}}
\put(1462,563){\blacken\ellipse{108}{108}}
\put(1462,563){\ellipse{108}{108}}
\put(1350,0){\makebox(0,0)[lb]{\smash{{{\SetFigFont{9}{10.8}{\rmdefault}{\mddefault}{\updefault}A}}}}}
\put(0,450){\makebox(0,0)[lb]{\smash{{{\SetFigFont{9}{10.8}{\rmdefault}{\mddefault}{\updefault}$G_1$}}}}}
\put(2550,450){\makebox(0,0)[lb]{\smash{{{\SetFigFont{9}{10.8}{\rmdefault}{\mddefault}{\updefault}$G_2$}}}}}
\end{picture}
}

%% file: eyeglass.eepic
\setlength{\unitlength}{0.00041667in}
\begingroup\makeatletter\ifx\SetFigFont\undefined%
\gdef\SetFigFont#1#2#3#4#5{%
  \reset@font\fontsize{#1}{#2pt}%
  \fontfamily{#3}\fontseries{#4}\fontshape{#5}%
  \selectfont}%
\fi\endgroup%
{\renewcommand{\dashlinestretch}{30}
\begin{picture}(4276,1097)(0,-10)
\put(937,541){\ellipse{1068}{1068}}
\put(3037,541){\ellipse{1068}{1068}}
\put(1462,541){\blacken\ellipse{108}{108}}
\put(1462,541){\ellipse{108}{108}}
\put(2512,541){\blacken\ellipse{108}{108}}
\put(2512,541){\ellipse{108}{108}}
\path(2100,503)(2550,503)
\path(1500,503)(2100,503)
\thicklines
\path(1920.000,443.000)(2100.000,503.000)(1920.000,563.000)
\put(1425,203){\makebox(0,0)[lb]{\smash{{{\SetFigFont{9}{10.8}{\rmdefault}{\mddefault}{\updefault}A}}}}}
\put(2325,203){\makebox(0,0)[lb]{\smash{{{\SetFigFont{9}{10.8}{\rmdefault}{\mddefault}{\updefault}B}}}}}
\put(1725,728){\makebox(0,0)[lb]{\smash{{{\SetFigFont{9}{10.8}{\rmdefault}{\mddefault}{\updefault}$g_3$}}}}}
\put(0,503){\makebox(0,0)[lb]{\smash{{{\SetFigFont{9}{10.8}{\rmdefault}{\mddefault}{\updefault}$g_1$}}}}}
\put(3600,503){\makebox(0,0)[lb]{\smash{{{\SetFigFont{9}{10.8}{\rmdefault}{\mddefault}{\updefault}$g_2$}}}}}
\end{picture}
}